\documentclass[aip,adv,reprint,floatfix]{revtex4-2}

\usepackage{syntonly}

\usepackage[utf8]{inputenc}

\usepackage[english]{babel}

\usepackage{xcolor}
\definecolor{lightblue}{RGB}{166,206,227}
\definecolor{darkblue}{RGB}{31,120,180}
\definecolor{lightgreen}{RGB}{178,223,138}
\definecolor{darkgreen}{RGB}{51,160,44}
\definecolor{teal}{RGB}{27,158,119}
\definecolor{orange}{RGB}{217,95,2}
\definecolor{purple}{RGB}{117,112,179}
\definecolor{magenta}{RGB}{231,41,138}

\usepackage{booktabs} 
\usepackage{siunitx} 
\usepackage{amsmath,amssymb,commath}

\usepackage{graphicx}

\usepackage[color = orange!30!white]{todonotes} 
\presetkeys{todonotes}{inline}{}

\newcommand*{\note}[1]{}
\newcommand*{\knnote}[1]{}

\begin{document}

\author{Tina Hecksher}
\email{tihe@ruc.dk}
\author{Kristine Niss}
\email{kniss@ruc.dk}
\affiliation{Glass and Time, IMFUFA, Department of Science and Environment, Roskilde University, Roskilde, Denmark}

\title{Single parameter aging and density scaling}
\date{\today}

\begin{abstract}

In a recent paper, Di Lisio \textit{et al.} [J.\ Chem.\ Phys.\ \textbf{159} 064505 (2023)] analyzed a series of temperature down-jumps using the single-parameter aging (SPA) ansatz combined with a specific assumption about density scaling in the out-of-equilibrium system and did not find a good prediction for the largest down-jumps.
In this paper we show that SPA in its original form does work for all their data including large jumps of $\Delta T>20$~K. Furthermore, we discuss different approaches to the extension of the density scaling concept to out-of-equilibrium systems.
\end{abstract}

\maketitle

\section{Introduction}
Physical aging is the gradual change in the physical properties of a glass over time, typically at temperatures around or just below its glass transition temperature, $T_g$. These changes in properties over time happen because the glass is out of equilibrium and spontaneously and irreversibly evolves towards the equilibrium state. 

Physical aging of glasses is an old topic dating back to at least Tool \& Eichlin in 1931\cite{Tool1931} rationalizing cooling and heating curves for oxide glasses. Nevertheless, the topic is far from understood and research in the area is active to this day, with a lot of important new work and methods appearing in the last couple of years\cite{Song2022,riechers2022, aguilera2022,douglass2022,mehri2022, vila2023,ruiz2023,dilisio2023,malek2023,henot2023,moch2023,herrero2023,gabriel2023_comparing,richert2023,henot2024,lancelotti2024,Moch2024,Bohmer2024,riechers2024}.

Experimental protocols for studying physical aging typically involve either temperature ramps or temperature jumps. The aging response to a given temperature input can be monitored by any experimental observable that can be acquired fast enough that the glass does not significantly change its properties during the measurement. Traditionally quantities such as refractive index \cite{Macedo1967,Huang2006}, volume \cite{Santore1991,Struik1997,Cangialosi2013}, enthalpy \cite{Moynihan1976,Kovacs1979}, mechanical creep or modulus \cite{Kovacs1963_dynamic,Echeverria1995,Soloukhin2003} have been used to quantify the out-of-equilibrium state of an aging glass. More recently gas permeability\cite{Huang2004}, dielectric susceptibility \cite{leheny1998,lunkenheimer2005,Richert2013,riechers2022}, and dynamic light scattering \cite{Bohmer2024} have been added to list.

Physical aging has obvious implications for the performance and properties of glassy materials used in applications, such as precision molding \cite{Dambon2009,Ming2020,Jiang2022,Vu2022}, oxide glasses used for optical fibers\cite{Jha2012} and displays \cite{Ellison2010}, plastics \cite{Weyhe2023}, polymers used for membranes\cite{Huang2004,Merrick2020}, bioactive glasses \cite{Boccaccini2010} etc., but aging also poses interesting questions for fundamental research. An intrinsic feature of aging is that it is non-linear and a central fundamental question is whether this non-linear out-of-equilibrium relaxation is governed by the same slow degrees of freedom that are responsible for the linear structural relaxation in the equilibrium state.
This has been the prevailing conception \cite{tool1946,narayanaswamy1971,Struik1977,Kovacs1979,Hutchinson1995,McKenna2020}, but it is possible that the physical aging mechanism could be governed by other processes. In particular, there are indications that for large up-jumps, aging proceeds in a heterogeneous way resembling a nucleation and growth rather than a gradual homogeneous softening \cite{herrero2023,vila2023,ruiz2023}. Furthermore, there is evidence that for large down-jumps, a faster aging mechanism sets in at short aging times \cite{Boucher2011,Eulatea2018,Song2022}, while the long time aging (i.e., close to equilibrium) remains dominated by the slow structural relaxation\cite{Cangialosi2013b}.

Altogether, these results suggests that other mechanism could be playing a role in aging, at least far from equilibrium. However, the classical models that connect aging directly to equilibrium linear relaxation are still widely used also  in applied glass-science e.g. Refs.~ \onlinecite{Huang2004, Dambon2009, Ming2020, Jiang2022, Vu2022, Weyhe2023}. Moreover, we and our colleagues have recently demonstrated with direct tests that this view can be confirmed experimentally \cite{riechers2022} and in simulations \cite{mehri2021} -- at least for small temperature jumps. 

\emph{If} (or when) the structural relaxation governs physical aging, then the out-of-equilibrium situation must be strongly connected to the properties that govern the equilibrium structural relaxation. In general, there is no consensus as to what controls the structural relaxation time, but it has been established that many organic systems obey so-called density scaling \cite{Alba2002,Tarjus2004,Casalini2004,Roland2005}. This means that the relaxation rate depends on a scaling parameter $\Gamma$, which in most cases is given by $\Gamma=\rho^\gamma/T$. For the systems where density scaling applies in equilibrium, it is natural to attempt to generalize the concept to out of equilibrium. One of us has earlier suggested how to do this \cite{niss2017,niss2022} and recently Di Lisio \textit{et al.} \cite{dilisio2023} suggested an alternative generalization. 

Single parameter aging (SPA)\cite{hecksher2015} is framework for describing and predicting aging within the class of models that directly connect aging to the equilibrium relaxation. Di Lisio \textit{et al.} \cite{dilisio2023} analyzed a series of temperature down-jumps using the SPA ansatz combined with assuming that density scaling as found in equilibrium for some systems generalizes directly to the out-of-equilibrium system. 
They conclude that SPA does not provide a good description for the largest down-jumps and use this result to argue that aging in those cases must proceed via a different -- faster --  mechanism than the alpha relaxation.
In this paper we show that SPA in its original form does work for all their data and argue that the reason for their finding could well be that their extension of the density scaling concept to out-of-equilibrium systems is not valid.

The paper is structured as follows: Section \ref{sec:spa} gives a short introduction to the concepts of material time and single parameter aging (SPA). Section \ref{sec:Tjumps} focuses on a comparison of simple linear SPA analysis and the version of SPA suggested by Di Lisio \textit{et al.} \cite{dilisio2023} applied to their temperature jump data. Section \ref{sec:densityscaling} compares the density scaling for out-of-equilibrium systems suggested by Di Lisio \textit{et al.} with the expansion of density scaling proposed by Niss \cite{niss2020,niss2022}. Our main points and results are summarized and discussed in Sec.~\ref{sec:conclusion}.

\section{Material time and single parameter aging}\label{sec:spa}

Aging is as mentioned a non-linear phenomenon in the sense that the response to a temperature step input depends on both the magnitude and the sign of the jump. A temperature up and down-jump of the same magnitude to the same target temperature will thus differ in shape: the down-jump will be stretched, while the up-jump is more compressed. This is called the asymmetry of approach \cite{tool1946,Kovacs1963_transition} and can be readily understood as a consequence of the change in relaxation time from initial state to the final state of the sample: In the up-jump, the material comes from a state of long relaxation time and adjusts gradually to a shorter relaxation time -- the dynamics in the material is ``auto-accelerated''. The opposite is true for the down-jump, where the dynamics of the material gradually becomes slower -- it is ``auto-retarded''\cite{mckenna1995,mckenna2012}. 

The Narayanaswamy model\cite{narayanaswamy1971} proposes that linearity of aging is restored when measured with respect to the ``internal clock'' of the material, the so-called material time $\xi$, meaning that the evolution of a measured property $X$ as a function of a temperature input $T(t)$ can be obtained by a linear convolution integral in the material time
\begin{equation}
    X(\xi)-X_\text{eq}(T) = -\alpha_X\int_{-\infty}^\xi M(\xi-\xi')\dot{T}\,d\xi\,,
\end{equation}
where $\dot{T}$ is the (material) time derivative of temperature, $\alpha_X$ is the thermal coefficient of the measured property (assumed to be constant), and $M$ is the memory kernel.
In his model, material time is related to the laboratory time through the instantaneous out-of-equilibrium relaxation time $\tau(t)$
\begin{equation}\label{eq:mattime}
    d\xi=\frac{dt}{\tau(t)}\,.
\end{equation}
In equilibrium, $\tau$ is the structural relaxation time given by the temperature and does not vary as a function of time. 
Out of equilibrium, it is assumed that one extra parameter, the fictive temperature $T_f$, is sufficient to describe the state of the system. This approach is often referred to as the Tool-Narayanaswamy-Moynihan (TNM) formalism\cite{tool1946,narayanaswamy1971,Moynihan1976b}.

In general, making predictions with the TNM formalism thus requires analytical expressions for the memory kernel, $M$, and for the out-of-equilibrium relaxation time, $\tau(T,T_f)$. Traditionally, non-exponential fitting functions have been proposed (such as the stretched exponential) for $M$ and various empirical expression for the out-of-equilibrium relaxation time, the most used being Moynihan's \cite{Moynihan1976b}. However, even \emph{in equilibrium}, we do not know the exact shape of $M$ or how $\tau$ behaves as a function of temperature (in fact these are the two big questions in glass science), so this introduces as a minimum 4 different fitting parameters depending on the choice of fitting functions.

Single parameter aging (SPA)\cite{hecksher2015} is a strategy to avoid fitting functions in the description of temperature jump experiments, but in its essence not different from the TNM formalism. The fundamental idea of SPA is that all properties of the aging material are governed by a single non-equilibrium parameter. It follows that there is a one-to-one connection between any two aging properties, including the instantaneous relaxation time itself and some measured property, $X$. 
The simplest assumption for this relation is linearity, i.e., that logarithm of the out-of-equilibrium relaxation time and the measured property are linearly connected, $\ln \tau(t) - \ln \tau_\text{eq} = a (X(t)-X_\text{eq})= a\Delta X(t)$. Assuming furthermore that the total change in the measured property is linear with changes in temperature, $\Delta X(0)=b\Delta T$, i.e., the difference between the equilibrium value at the initial temperature before the jump and the equilibrium value at the new temperature, lead to the following expression for the out-of-equilibrium relaxation time 
\begin{equation}\label{eq:gamma}
    \ln \tau(t)= \ln\tau_\text{eq} + g(R)\,,
\end{equation}
with
\begin{equation}
    g(R)=c\Delta T R(t)
\end{equation}
where $c=ab$ and $R(t)=\frac{\Delta X(t)}{\Delta X(0)}$ is the normalized relaxation function. Thus, the out-of-equilibrium relaxation time, $\ln\tau(t)$, is given by the equilibrium relaxation time at the annealing temperature, $\ln\tau_{\text{eq}}$, plus a function of the normalized relaxation, $g(R)=c\Delta TR$.

If TNM formalism applies, the relaxation curves from all jumps collapse when plotted as a function material time by a transformation of the time axis according to Eq.~(\ref{eq:mattime}). The transformation can also be reversed such that -- if the material time and the out-of-equilibrium relaxation time $\tau(t)$  corresponding to a given temperature jump is known -- we can obtain the laboratory time corresponding to that jump. In this sense, one can think of $R$ as the independent variable and time as the dependent variable. This is true for any temperature jump and consequently we can transform the time axis from one non-linear jump to another.
In a condensed form, the above can be written as 
\begin{equation}
\begin{split}\label{eq:transf}
    \frac{dt_1(R)}{\tau_1}&=d\xi(R)=\frac{dt_2(R)}{\tau_2} \Rightarrow \\ dt_2&(R)= \frac{\tau_2}{\tau_1}dt_1(R)\,,
\end{split}
\end{equation}
where the subscripts 1 and 2 refer to two different temperature jumps, 1 and 2.

SPA gives a relation between the out-of-equilibrium relaxation time and the measured quantity (Eq.~(\ref{eq:gamma})), which inserted in  Eq.~(\ref{eq:transf}) leads to the time axis for temperature jump 2
\begin{equation}\label{eq:tint}
t_2(R)=\frac{\tau_{\text{eq},2}}{\tau_{\text{eq},1}}\int_0^{t_1} \exp \left\{c(\Delta T_2-\Delta T_1)R  \right\}\, dt_1(R)\,.
\end{equation}

Thus, given $c$ and the equilibrium relaxation times, $\tau_{\text{eq},n}$, the SPA provides a recipe for obtaining any temperature-jump relaxation curve from a given measured temperature jump. 

The linearity assumptions in the SPA approach reduce the number of fitting parameters to one ($c$), but are not strictly required for single parameter aging to hold. 

\section{SPA analysis of temperature jumps}\label{sec:Tjumps}

SPA in its simplest form described above has been demonstrated to work for small temperature jumps \cite{hecksher2015,roed2019,hecksher2019,riechers2022}. For large temperature jumps neither of the two linearity assumptions necessarily hold and this simplest formulation of SPA may well break down and/or other aging mechanism may set in. Indeed, there are indications that SPA breaks down for larger jumps, in particular larger up-jumps \cite{roed2019,riechers2022}. 

Di Lisio et al. measured a series of temperature down-jumps of sizes up to 22.5 K using fast scanning calorimetry for five different organic liquids: phenolphthalein dimethyl ether (PDE), o-Cresolphthalein dimethyl ether (KDE), 
 o-terphenyl (OTP), bisphenol-C-dimethylether (BMPC), and 1,1-bis (4-methoxyphenyl)cyclohexane (BMMPC). They used the enthalpy overshoot in heating to extract the fictive temperature, $T_f$, after a different waiting times at the annealing temperature, $T_a$. In this way they constructed curves for the temporal evolution of $T_f$ during aging. Subsequently, these curves were analyzed in the SPA framework combined with a density scaling assumption. 

It is by no means surprising if SPA did not work for the large temperature down-jumps presented by Di Lisio \textit{et al}. However, applying SPA in this simplest form, i.e. with linearity assumptions and $c$ as a sole fitting parameter for all curves, we demonstrate below that SPA does capture all of the measured data, even the largest down-jumps reported by Di Lisio \textit{et al}.

\begin{figure*}
    \centering
    \includegraphics[width=\textwidth]{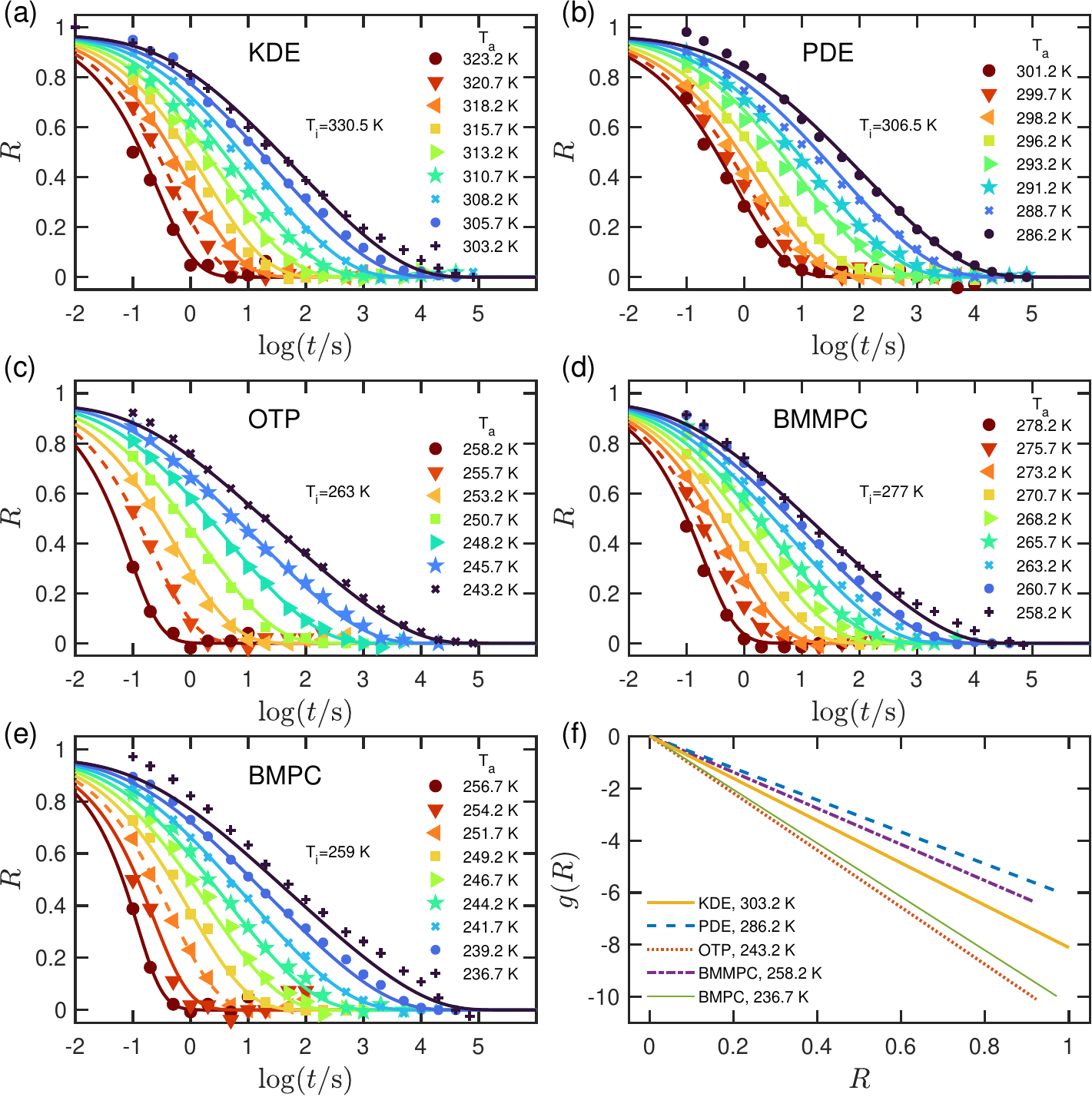}
    \caption{The simple SPA analysis performed on the data from Di Lisio \cite{dilisio2023} for (a) KDE, (b) PDE, (c) OTP, (d) BMMPC, and (e) BMPC. Symbols are the measured data points, full lines the predicted aging curves. The curves predicted from SPA are based on the dashed line, which is a stretched exponential fit to the same-colored data points (see also Supplemental Material). (f) Shows the simple SPA expression for $g(R)=c\Delta T R$ as a function of $R$ for the largest temperature jump for each of the liquids.}
    \label{fig:org_spa}
\end{figure*}

\begin{figure*}
    \centering
    \includegraphics[width=\textwidth]{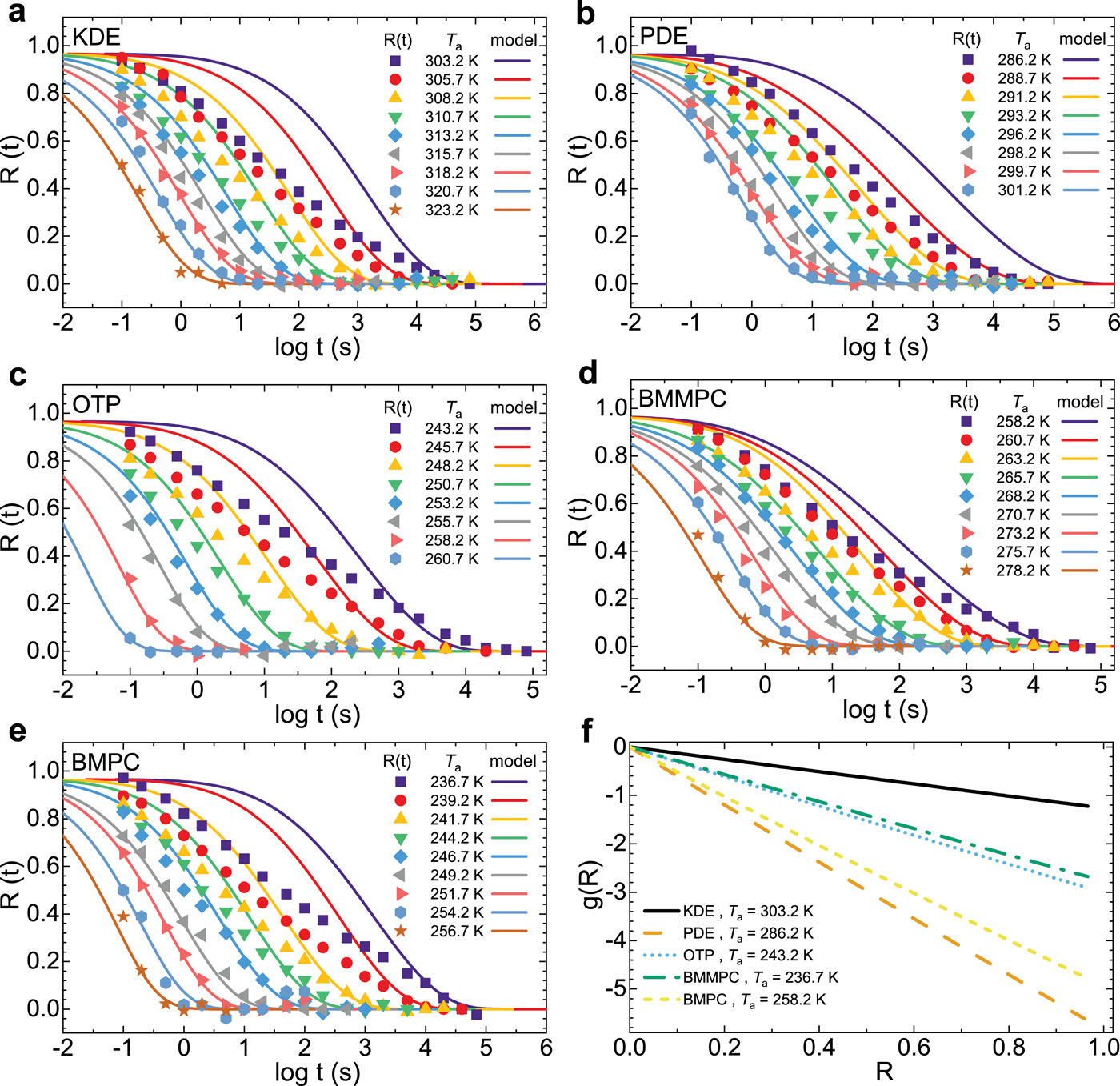}
    \caption{Data and model curves for the same data as in Fig.~\ref{fig:org_spa}. The model curves are based on a density scaling assumption for the out-of-equilibrium system. Reproduced from Di Lisio \textit{et al} \cite{dilisio2023}.\\ \note{This is an open access article distributed under the terms of the Creative Commons CC BY license, which permits unrestricted use, distribution, and reproduction in any medium, provided the original work is properly cited.\\ You are not required to obtain permission to reuse this article.}}
    \label{fig:dens_spa}
\end{figure*}

Figure \ref{fig:org_spa}(A-E) shows the resulting SPA analysis (see Supplemental Material for details of the analysis). For each liquid, one jump was selected as the predictor of the rest. Since the data are sparse, we fitted the selected jump with a stretched exponential to serve as interpolation between data points. 
The curve fitted to the chosen jump is shown as a dashed line. This curve is then used to generate the other temperature jumps from Eq.~(\ref{eq:tint}) and the results are shown as full lines. For all data sets and all temperatures, there is a decent agreement between the predicted curve and and the measured data points. The fitted values for $c$ is given in Table \ref{tab:cvalues}.
\begin{table}[]
    \centering
    \begin{tabular}{l|c}
    \hline
    \hline
    Liquid & $c$ (K$^{-1}$) \\
    \hline
       KDE  &  0.30 \\
       PDE  &  0.30 \\
       OTP  &  0.55 \\
       BMMPC  &  0.37 \\
       BMPC  & 0.46 \\
    \hline
    \hline
    \end{tabular}
    \caption{Fitted values for $c$ of Eq.~(\ref{eq:tint}).}
    \label{tab:cvalues}
\end{table}
There are small deviations from the predictions for the largest temperature jumps for KDE, BMMPC, and BMPC at long times. For BMPC the predicted curve for the largest temperature jump is consistently below the data points. However, this is a simple scaling and can be explained by uncertainty in determining the equilibrium relaxation time from the data (see Supplementary Material for more details). 
Overall, SPA works to a remarkably good degree, even for the largest temperature jumps of $\Delta T>20$~K, when conducted in its original and simplest form.
Figure \ref{fig:org_spa}(f) shows $g(R)$ as a function of $R$ for the largest jumps for each liquid. The simple SPA $g(R)$ is linear by definition.

The analysis by Di Lisio \textit{et al.} \cite{dilisio2023} is shown in Fig.~\ref{fig:dens_spa} for comparison. Their predicted model curves deviate significantly from the data, especially for the larger down-jumps. These model curves are based a SPA analysis similar to Eq.~(\ref{eq:gamma}), however using a different expression for $g(R)$,
\begin{equation}\label{eq:gR_dilisio}
    g(R)= \frac{C}{T_fV_0^{\gamma}}\left( \frac{1}{(1-\Delta T_f (\alpha_l-\Delta \alpha R(t)))^{\gamma}}-\frac{1}{(1 - \alpha_l \Delta T_f )^{\gamma}} \right)\,,
\end{equation}
where $C$ is a constant and $\gamma$ is the density scaling exponent (both determined by equilibrium measurements), $\alpha_l$ is the thermal expansivity of the liquid, $\Delta \alpha$ the difference in thermal expansivity between liquid and glass, and $V_0$ the initial specific volume.

Their expression is based on three assumptions: 
1) that volume and fictive temperature relax in the same way so the that the instantaneous volume can be calculated as $V(T_a,t) = V_0 \left[1 - \alpha_l( T_f (0) - T_f (t)) - \alpha_g (T_f (t)- T_a )\right]$, where $\alpha_l$ and $\alpha_g$ are the thermal expansion coefficients of the liquid and the glass,  2) that the instantaneous relaxation time is given by a direct use of density scaling for equilibrium to the out-of-equilibrium situation $\tau(T_a,t) = F\left( \frac{C}{T_aV(T_a,t)^\gamma} \right)$, and 3) that the function $F$ is an exponential. 

Figure \ref{fig:dens_spa}(f) shows that the $g(R)$ from Eq.~(\ref{eq:gR_dilisio}) with the inserted parameters is virtually linear in $R$ in the relevant range despite the complexity of the expression. This means that, effectively Di Lisio \textit{et al.} have fixed $c$ in the simple SPA analysis (instead of keeping it as a fitting parameter) by introducing a series of assumption for the out-of-equilibrium relaxation time.
Thus, in the approach of Di Lisio \textit{et al.} does not directly test SPA because it requires that also the additional assumptions for the out-of-equilibrium relaxation time to hold. 
The linearity of Di Lisio's $g(R)$ and the fact that another linear $g(R)=c\Delta T R$ (given by the simplest SPA with a single value of $c$) does accurately capture the aging curves, suggests that one (or more) of the assumptions made by Di Lisio \textit{et al.} for the out-of-equilibrium relaxation time is the reason for the failure of their SPA analysis. Below we will discuss their density scaling assumption and compare it to the generalized density scaling suggested in Ref.~\onlinecite{niss2022}.

\section{Combining single parameter aging with density scaling}\label{sec:densityscaling}
While the mechanism and functional forms are unknown, the relaxation time in equilibrium is of course given by the equilibrium thermodynamics state of the system. Generally the state of an equilibrium system is uniquely given by two thermodynamic state variables. It is most common to describe the system by pressure and temperature or density and temperature, but it could just as well be for example pressure and density or density and entropy. The point is that there are only two independent state variables and that the relaxation time as well as all other properties of the equilibrium liquid depend on these. However, for an out-of-equilibrium system this is not the case; at least one parameter has to be added to account for the departure from equilibrium. Single parameter aging refers to a situation where the out-of-equilibrium relaxation time during aging is controlled by two thermodynamic variables plus a single additional parameter. 

In the classical TNM-model the parameter is, as mentioned in Sec. \ref{sec:spa}, the fictive temperature, $T_f$. The state of the system is then given by $T_f$ and the actual temperature $T$. In this classical work it is implicitly understood that the sample is kept under ambient isobaric conditions, which means that also in this case two state variables plus a parameter describe the system. Another way of describing the out-of-equilibrium state by three parameters is to acknowledge that temperature, density and pressure are three independent variables and if single parameter aging holds, then these can be used to fully characterize the system state, as developed in Ref.~\onlinecite{hecksher2019}.

In equilibrium it has been established that many systems obey density scaling, which means that the relaxation depends on a scaling parameter $\Gamma$, which is given by $\Gamma=e(\rho)/T$. In many cases power law density scaling works well, which means that $e(\rho)=\rho^\gamma$ and $\Gamma=\rho^\gamma/T$. This means that the relaxation time depends on one thermodynamic state variable rather than two: $\tau_\text{eq}=f(\Gamma)$. 

The density scaling ansatz by Di Lisio \textit{et al.} is a direct use of power-law density scaling out of equilibrium. Thus they assume that $\Gamma=\rho^\gamma/T$ controls the aging rate out of equilibrium exactly as it does in equilibrium. In other words, only one thermodynamic parameter describes the system even when it is out of equilibrium, while there is no out-of-equilibrium parameter. If this anzats holds, $\tau$ cannot age under isochoric isothermic condition (e.g., after a temperature jump at constant volume) and aging of other properties would be linear. However, it is very well established from computer simulations that $\tau$ does change under isochoric conditions \cite{Castillo2007, mehri2021} and that isochoric aging is nonlinear \cite{DiLeonardo2000, Gnan2010, Schober2012, herrero2023}. It is theoretically possible that Di Lisio's ansatz could hold under isobaric conditions, meaning that it should be understood that the rate is governed by $\Gamma=\rho^\gamma/T$ \emph{and} the pressure. However, it is a bit counter-intuitive to generalize density scaling to something that only works at ambient pressure. Moreover, the finding of Di Lisio \textit{et al.}, that they cannot describe the aging curves with their model, while the original SPA does describe the aging curves, suggest that it may be the density scaling anzats which does not hold.  

In 2022 one of us proposed another way to generalize density scaling to aging systems \cite{niss2022}. Here the idea is that for systems with density scaling in equilibrium one thermodynamic variable, plus an out-of-equilibrium parameter, are needed to describe the system during aging. The suggestion is that the relevant thermodynamic variable is $\Gamma$ defined as in equilibrium and given uniquely by density and temperature ($\Gamma=e(\rho)/T$). The departure from equilibrium is proposed to be described by the parameter $\Gamma_{fic}$, which is defined to approach $\Gamma$ as the system approaches equilibrium. 

To get an expression for the relaxation time during aging in this framework, the starting point is the equilibrium density scaling in the original version of Alba-Simionesco, Tarjus, and Kivelson \cite{Alba2002,Tarjus2004}, where $\Gamma$ is defined as $\Gamma=e(\rho)/T$ with $e(\rho)$ interpreted as an energy scale that only depends on density while the topology of the energy landscape depends on both temperature and density through $\Gamma$. This leads to the following expression for the activation energy in equilibrium:
\begin{equation}
    \Delta E_\text{eq}=e(\rho)F\left( \frac{e(\rho)}{T}\right) =e(\rho) F(\Gamma)
\end{equation}
and the expression for the relaxation time becomes
\begin{equation}
    \tau_\text{eq}=\tau_0 \text{exp}\left( \frac{e(\rho)}{T}F\left( \frac{e(\rho)}{T}\right) \right)=\tau_0 \text{exp}\left( \Gamma F( \Gamma ) \right)
\end{equation}

When generalizing to out-of-equilibrium situation we assume that the energy scale $e(\rho)$ is controlled by density while the topology of the energy landscape is controlled by the non-equilibrium structural state of the system, described by the paramter, $\Gamma_{fic}$. This parameter can be expressed as $\Gamma_{fic}=e(\rho)/T_{\Gamma f}$, where $e(\rho)$ is defined as in equilibrium and $T_{\Gamma f}$ is the $\Gamma_{fic}$-temperature ({Gamma-fictive  temperature}), which is conceptually similar to the
classical fictive temperature, though it has a quite different behavior \cite{niss2022}.

In this framework the activation energy during aging becomes
\begin{equation}
    \Delta E_\text{aging}=e(\rho)F(\Gamma_{fic}) 
\end{equation}
and it follows that the relaxation time during aging should be given by
\begin{equation}
\tau_{\text{aging}}=\tau_0 \text{exp}\left( \Gamma F( \Gamma_{fic} ) \right)\label{eq:KNscaling}
\end{equation}
where the functional form of $F$ is the same as in equilibrium. 

The ansatz in this framework is that $\Gamma_{fic}$ which is connected to the microscopic structure, needs time to change. Right after a temperature jump it will therefore be equal to the value it had before the jump was performed. Thus, Eq.~(\ref{eq:KNscaling}) predicts that the relaxation time is further from equilibrium at short times than predicted by Di Lisio \textit{et al.}. In the case of down-jumps in temperature it means that the  curves predicted by Eq.~(\ref{eq:KNscaling}) will be more stretched than those predicted by Di Lisio \textit{et al.} consistent with the data in Fig.~\ref{fig:dens_spa} being more stretched than the prediction.  Hénot \textit{et al.} \cite{henot2024} have earlier this year made a similar qualitative comparison and likewise find that Eq.~(\ref{eq:KNscaling}) has a better chance of representing the data.  Hénot \textit{et al.} consider up-jumps, and in this case it means that Eq.~(\ref{eq:KNscaling}) predicts longer relaxation times right after the temperature jump is performed than predicted by the Di Lisio expression, leading to a more compressed curve. 

The down side of the expression in Eq.~(\ref{eq:KNscaling}) is that actual quantitative tests are difficult and have not yet been performed. In order to do that, it is necessary to carry out dedicated high pressure aging experiments as proposed in Ref.~\onlinecite{niss2022} and preferably to have samples where the functional forms of $F(\Gamma)$ and $e(\rho)$ are known from equilibrium in addition to knowing both density and relaxation time during aging. 

\section{Summary and concluding remarks}\label{sec:conclusion}

We have shown that the temperature jump data of Di Lisio \text{et al} \cite{dilisio2023} can be described fully by the single parameter ansatz with one free fitting parameter for each sample. Di Lisio \textit{et al.} arrived at the opposite conclusion. However, their conclusion is based on the SPA ansatz \emph{and} assumptions for density scaling in the out-of-equilibrium system. At this point it is not known if density scaling can be generalized to aging nor how to do it, but we argue that the following logic should apply: In general, two variables determine the equilibrium state of a liquid (for example density and temperature). During aging at least one extra parameter is added to account for the departure from equilibrium. In other words, three variables determine the non-equilibrium state if single parameter aging holds (the word \emph{single} refers to the fact that only one parameter describes the departure from equilibrium). For the class of liquids where density scaling holds one thermodynamic variable determines the relaxation time in equilibrium. As for the general state a single additional parameter is needed to describe the departure from equilibrium during aging. Generalization of density scaling to aging is thus expected to give an expression where the relaxation time is given by two variables.    

We cannot rule out that the assumptions for out-of-equilibrium density scaling of Di Lisio \textit{et al.} are correct and that SPA does not work for large temperature jumps. In fact, the latter is very likely true; in particular for large up-jumps. We make the point that it is not trivial how to generalize density scaling, so
when testing both SPA and out-of-equilibrium density scaling \emph{at the same time}, it is not possible to determine which of the two is wrong, if the test fails. A generalization of density scaling to the out-of-equilibrium situation is intriguing and possibly a fruitful path, however more quantitative studies are necessary before we know which generalization scenario is correct.

\section*{Acknowledgements}
We thank Jeppe C.\ Dyre and Nicholas P.\ Bailey for discussions of the manuscript. Furthermore, we are grateful for the continued open discussions with Daniele Cangialosi about aging in general and SPA in particular. Finally, we would like to thank Valerio Di Lisio and Daniele Cangialosi for generously sharing their data. This work was supported by the VILLUM Foundation's Matter Grant No. VIL16515.

\section*{Data Availability}
Data sharing is not applicable to this article as no new data were
created or analyzed in this study.
Data presented in the paper are from Ref.~\onlinecite{dilisio2023}


\begin{thebibliography}{71}%
\makeatletter
\providecommand \@ifxundefined [1]{%
 \@ifx{#1\undefined}
}%
\providecommand \@ifnum [1]{%
 \ifnum #1\expandafter \@firstoftwo
 \else \expandafter \@secondoftwo
 \fi
}%
\providecommand \@ifx [1]{%
 \ifx #1\expandafter \@firstoftwo
 \else \expandafter \@secondoftwo
 \fi
}%
\providecommand \natexlab [1]{#1}%
\providecommand \enquote  [1]{``#1''}%
\providecommand \bibnamefont  [1]{#1}%
\providecommand \bibfnamefont [1]{#1}%
\providecommand \citenamefont [1]{#1}%
\providecommand \href@noop [0]{\@secondoftwo}%
\providecommand \href [0]{\begingroup \@sanitize@url \@href}%
\providecommand \@href[1]{\@@startlink{#1}\@@href}%
\providecommand \@@href[1]{\endgroup#1\@@endlink}%
\providecommand \@sanitize@url [0]{\catcode `\\12\catcode `\$12\catcode
  `\&12\catcode `\#12\catcode `\^12\catcode `\_12\catcode `\%12\relax}%
\providecommand \@@startlink[1]{}%
\providecommand \@@endlink[0]{}%
\providecommand \url  [0]{\begingroup\@sanitize@url \@url }%
\providecommand \@url [1]{\endgroup\@href {#1}{\urlprefix }}%
\providecommand \urlprefix  [0]{URL }%
\providecommand \Eprint [0]{\href }%
\providecommand \doibase [0]{https://doi.org/}%
\providecommand \selectlanguage [0]{\@gobble}%
\providecommand \bibinfo  [0]{\@secondoftwo}%
\providecommand \bibfield  [0]{\@secondoftwo}%
\providecommand \translation [1]{[#1]}%
\providecommand \BibitemOpen [0]{}%
\providecommand \bibitemStop [0]{}%
\providecommand \bibitemNoStop [0]{.\EOS\space}%
\providecommand \EOS [0]{\spacefactor3000\relax}%
\providecommand \BibitemShut  [1]{\csname bibitem#1\endcsname}%
\let\auto@bib@innerbib\@empty
\bibitem [{\citenamefont {Tool}\ and\ \citenamefont
  {Eichlin}(1931)}]{Tool1931}%
  \BibitemOpen
  \bibfield  {author} {\bibinfo {author} {\bibfnamefont {A.~Q.}\ \bibnamefont
  {Tool}}\ and\ \bibinfo {author} {\bibfnamefont {C.~G.}\ \bibnamefont
  {Eichlin}},\ }\bibfield  {title} {\enquote {\bibinfo {title} {Variations
  caused in the heating curves of glass by heat treatment},}\ }\href
  {https://doi.org/https://doi.org/10.1111/j.1151-2916.1931.tb16602.x}
  {\bibfield  {journal} {\bibinfo  {journal} {Journal of the American Ceramic
  Society}\ }\textbf {\bibinfo {volume} {14}},\ \bibinfo {pages} {276--308}
  (\bibinfo {year} {1931})}\BibitemShut {NoStop}%
\bibitem [{\citenamefont {Song}\ \emph {et~al.}(2022)\citenamefont {Song},
  \citenamefont {Rodríguez-Tinoco}, \citenamefont {Mathew},\ and\
  \citenamefont {Napolitano}}]{Song2022}%
  \BibitemOpen
  \bibfield  {author} {\bibinfo {author} {\bibfnamefont {Z.}~\bibnamefont
  {Song}}, \bibinfo {author} {\bibfnamefont {C.}~\bibnamefont
  {Rodríguez-Tinoco}}, \bibinfo {author} {\bibfnamefont {A.}~\bibnamefont
  {Mathew}},\ and\ \bibinfo {author} {\bibfnamefont {S.}~\bibnamefont
  {Napolitano}},\ }\bibfield  {title} {\enquote {\bibinfo {title} {Fast
  equilibration mechanisms in disordered materials mediated by slow liquid
  dynamics},}\ }\href {https://doi.org/DOI: 10.1126/sciadv.abm7154} {\bibfield
  {journal} {\bibinfo  {journal} {Science Advances}\ }\textbf {\bibinfo
  {volume} {8}},\ \bibinfo {pages} {eabm7154} (\bibinfo {year}
  {2022})}\BibitemShut {NoStop}%
\bibitem [{\citenamefont {Riechers}\ \emph {et~al.}(2022)\citenamefont
  {Riechers}, \citenamefont {Roed}, \citenamefont {Mehri}, \citenamefont
  {Ingebrigtsen}, \citenamefont {Hecksher}, \citenamefont {Dyre},\ and\
  \citenamefont {Niss}}]{riechers2022}%
  \BibitemOpen
  \bibfield  {author} {\bibinfo {author} {\bibfnamefont {B.}~\bibnamefont
  {Riechers}}, \bibinfo {author} {\bibfnamefont {L.~A.}\ \bibnamefont {Roed}},
  \bibinfo {author} {\bibfnamefont {S.}~\bibnamefont {Mehri}}, \bibinfo
  {author} {\bibfnamefont {T.~S.}\ \bibnamefont {Ingebrigtsen}}, \bibinfo
  {author} {\bibfnamefont {T.}~\bibnamefont {Hecksher}}, \bibinfo {author}
  {\bibfnamefont {J.~C.}\ \bibnamefont {Dyre}},\ and\ \bibinfo {author}
  {\bibfnamefont {K.}~\bibnamefont {Niss}},\ }\bibfield  {title} {\enquote
  {\bibinfo {title} {Predicting nonlinear physical aging of glasses from
  equilibrium relaxation via the material time},}\ }\href@noop {} {\bibfield
  {journal} {\bibinfo  {journal} {Sci. Adv.}\ }\textbf {\bibinfo {volume}
  {8}},\ \bibinfo {pages} {eabl9809} (\bibinfo {year} {2022})}\BibitemShut
  {NoStop}%
\bibitem [{\citenamefont {Elizondo-Aguilera}, \citenamefont {Rizzo},\ and\
  \citenamefont {Voigtmann}(2022)}]{aguilera2022}%
  \BibitemOpen
  \bibfield  {author} {\bibinfo {author} {\bibfnamefont {L.~F.}\ \bibnamefont
  {Elizondo-Aguilera}}, \bibinfo {author} {\bibfnamefont {T.}~\bibnamefont
  {Rizzo}},\ and\ \bibinfo {author} {\bibfnamefont {T.}~\bibnamefont
  {Voigtmann}},\ }\bibfield  {title} {\enquote {\bibinfo {title} {From subaging
  to hyperaging in structural glasses},}\ }\href
  {https://doi.org/10.1103/PhysRevLett.129.238003} {\bibfield  {journal}
  {\bibinfo  {journal} {Phys. Rev. Lett.}\ }\textbf {\bibinfo {volume} {129}},\
  \bibinfo {pages} {238003} (\bibinfo {year} {2022})}\BibitemShut {NoStop}%
\bibitem [{\citenamefont {Douglass}\ and\ \citenamefont
  {Dyre}(2022)}]{douglass2022}%
  \BibitemOpen
  \bibfield  {author} {\bibinfo {author} {\bibfnamefont {I.~M.}\ \bibnamefont
  {Douglass}}\ and\ \bibinfo {author} {\bibfnamefont {J.~C.}\ \bibnamefont
  {Dyre}},\ }\bibfield  {title} {\enquote {\bibinfo {title} {Distance-as-time
  in physical aging},}\ }\href {https://doi.org/10.1103/PhysRevE.106.054615}
  {\bibfield  {journal} {\bibinfo  {journal} {Phys. Rev. E}\ }\textbf {\bibinfo
  {volume} {106}},\ \bibinfo {pages} {054615} (\bibinfo {year}
  {2022})}\BibitemShut {NoStop}%
\bibitem [{\citenamefont {Mehri}, \citenamefont {Costigliola},\ and\
  \citenamefont {Dyre}(2022)}]{mehri2022}%
  \BibitemOpen
  \bibfield  {author} {\bibinfo {author} {\bibfnamefont {S.}~\bibnamefont
  {Mehri}}, \bibinfo {author} {\bibfnamefont {L.}~\bibnamefont {Costigliola}},\
  and\ \bibinfo {author} {\bibfnamefont {J.~C.}\ \bibnamefont {Dyre}},\
  }\bibfield  {title} {\enquote {\bibinfo {title} {Single-parameter aging in
  the weakly nonlinear limit},}\ }\href {https://doi.org/10.3390/thermo2030013}
  {\bibfield  {journal} {\bibinfo  {journal} {Thermo}\ }\textbf {\bibinfo
  {volume} {2}},\ \bibinfo {pages} {160--170} (\bibinfo {year}
  {2022})}\BibitemShut {NoStop}%
\bibitem [{\citenamefont {Vila-Costa}\ \emph {et~al.}(2023)\citenamefont
  {Vila-Costa}, \citenamefont {Gonzalez-Silveira}, \citenamefont
  {Rodríguez-Tinoco}, \citenamefont {Rodríguez-López},\ and\ \citenamefont
  {Rodriguez-Viejo}}]{vila2023}%
  \BibitemOpen
  \bibfield  {author} {\bibinfo {author} {\bibfnamefont {A.}~\bibnamefont
  {Vila-Costa}}, \bibinfo {author} {\bibfnamefont {M.}~\bibnamefont
  {Gonzalez-Silveira}}, \bibinfo {author} {\bibfnamefont {C.}~\bibnamefont
  {Rodríguez-Tinoco}}, \bibinfo {author} {\bibfnamefont {M.}~\bibnamefont
  {Rodríguez-López}},\ and\ \bibinfo {author} {\bibfnamefont
  {J.}~\bibnamefont {Rodriguez-Viejo}},\ }\bibfield  {title} {\enquote
  {\bibinfo {title} {Emergence of equilibrated liquid regions within the
  glass},}\ }\href@noop {} {\bibfield  {journal} {\bibinfo  {journal} {Nat.
  Phys.}\ }\textbf {\bibinfo {volume} {19}},\ \bibinfo {pages} {114} (\bibinfo
  {year} {2023})}\BibitemShut {NoStop}%
\bibitem [{\citenamefont {Ruiz-Ruiz}\ \emph {et~al.}(2023)\citenamefont
  {Ruiz-Ruiz}, \citenamefont {Vila-Costa}, \citenamefont {Bar}, \citenamefont
  {Rodríguez-Tinoco}, \citenamefont {Gonzalez-Silveira}, \citenamefont
  {Plaza}, \citenamefont {Alcalá}, \citenamefont {Fraxedas},\ and\
  \citenamefont {Rodriguez-Viejo}}]{ruiz2023}%
  \BibitemOpen
  \bibfield  {author} {\bibinfo {author} {\bibfnamefont {M.}~\bibnamefont
  {Ruiz-Ruiz}}, \bibinfo {author} {\bibfnamefont {A.}~\bibnamefont
  {Vila-Costa}}, \bibinfo {author} {\bibfnamefont {T.}~\bibnamefont {Bar}},
  \bibinfo {author} {\bibfnamefont {C.}~\bibnamefont {Rodríguez-Tinoco}},
  \bibinfo {author} {\bibfnamefont {M.}~\bibnamefont {Gonzalez-Silveira}},
  \bibinfo {author} {\bibfnamefont {J.~A.}\ \bibnamefont {Plaza}}, \bibinfo
  {author} {\bibfnamefont {J.}~\bibnamefont {Alcalá}}, \bibinfo {author}
  {\bibfnamefont {J.}~\bibnamefont {Fraxedas}},\ and\ \bibinfo {author}
  {\bibfnamefont {J.}~\bibnamefont {Rodriguez-Viejo}},\ }\bibfield  {title}
  {\enquote {\bibinfo {title} {Real-time microscopy of the relaxation of a
  glass},}\ }\href@noop {} {\bibfield  {journal} {\bibinfo  {journal} {Nat.
  Phys.}\ }\textbf {\bibinfo {volume} {19}},\ \bibinfo {pages} {1} (\bibinfo
  {year} {2023})}\BibitemShut {NoStop}%
\bibitem [{\citenamefont {Di~Lisio}, \citenamefont {Stavropoulou},\ and\
  \citenamefont {Cangialosi}(2023)}]{dilisio2023}%
  \BibitemOpen
  \bibfield  {author} {\bibinfo {author} {\bibfnamefont {V.}~\bibnamefont
  {Di~Lisio}}, \bibinfo {author} {\bibfnamefont {V.-M.}\ \bibnamefont
  {Stavropoulou}},\ and\ \bibinfo {author} {\bibfnamefont {D.}~\bibnamefont
  {Cangialosi}},\ }\bibfield  {title} {\enquote {\bibinfo {title} {Physical
  aging in molecular glasses beyond the $\alpha$ relaxation},}\ }\href@noop {}
  {\bibfield  {journal} {\bibinfo  {journal} {J. Chem. Phys.}\ }\textbf
  {\bibinfo {volume} {159}},\ \bibinfo {pages} {064505} (\bibinfo {year}
  {2023})}\BibitemShut {NoStop}%
\bibitem [{\citenamefont {Málek}(2023)}]{malek2023}%
  \BibitemOpen
  \bibfield  {author} {\bibinfo {author} {\bibfnamefont {J.}~\bibnamefont
  {Málek}},\ }\bibfield  {title} {\enquote {\bibinfo {title} {Structural
  relaxation rate and aging in amorphous solids},}\ }\href@noop {} {\bibfield
  {journal} {\bibinfo  {journal} {J. Phys. Chem. C}\ }\textbf {\bibinfo
  {volume} {127}},\ \bibinfo {pages} {6080} (\bibinfo {year}
  {2023})}\BibitemShut {NoStop}%
\bibitem [{\citenamefont {Hénot}\ and\ \citenamefont
  {Ladieu}(2023)}]{henot2023}%
  \BibitemOpen
  \bibfield  {author} {\bibinfo {author} {\bibfnamefont {M.}~\bibnamefont
  {Hénot}}\ and\ \bibinfo {author} {\bibfnamefont {F.}~\bibnamefont
  {Ladieu}},\ }\bibfield  {title} {\enquote {\bibinfo {title} {Non-linear
  physical aging of supercooled glycerol induced by large upward ideal
  temperature steps monitored through cooling experiments},}\ }\href@noop {}
  {\bibfield  {journal} {\bibinfo  {journal} {J. Chem. Phys.}\ }\textbf
  {\bibinfo {volume} {158}},\ \bibinfo {pages} {224504} (\bibinfo {year}
  {2023})}\BibitemShut {NoStop}%
\bibitem [{\citenamefont {Moch}, \citenamefont {Böhmer},\ and\ \citenamefont
  {Gainaru}(2023)}]{moch2023}%
  \BibitemOpen
  \bibfield  {author} {\bibinfo {author} {\bibfnamefont {K.}~\bibnamefont
  {Moch}}, \bibinfo {author} {\bibfnamefont {R.}~\bibnamefont {Böhmer}},\ and\
  \bibinfo {author} {\bibfnamefont {C.}~\bibnamefont {Gainaru}},\ }\bibfield
  {title} {\enquote {\bibinfo {title} {Temperature oscillations provide access
  to high-order physical aging harmonics of a glass forming melt},}\
  }\href@noop {} {\bibfield  {journal} {\bibinfo  {journal} {J. Chem. Phys.}\
  }\textbf {\bibinfo {volume} {159}},\ \bibinfo {pages} {221102} (\bibinfo
  {year} {2023})}\BibitemShut {NoStop}%
\bibitem [{\citenamefont {Herrero}\ \emph {et~al.}(2023)\citenamefont
  {Herrero}, \citenamefont {Scalliet}, \citenamefont {Ediger},\ and\
  \citenamefont {Berthier}}]{herrero2023}%
  \BibitemOpen
  \bibfield  {author} {\bibinfo {author} {\bibfnamefont {C.}~\bibnamefont
  {Herrero}}, \bibinfo {author} {\bibfnamefont {C.}~\bibnamefont {Scalliet}},
  \bibinfo {author} {\bibfnamefont {M.}~\bibnamefont {Ediger}},\ and\ \bibinfo
  {author} {\bibfnamefont {L.}~\bibnamefont {Berthier}},\ }\bibfield  {title}
  {\enquote {\bibinfo {title} {Two-step devitrification of ultrastable
  glasses},}\ }\href@noop {} {\bibfield  {journal} {\bibinfo  {journal} {Proc.
  Natl. Acad. Sci. U. S. A.}\ }\textbf {\bibinfo {volume} {120}},\ \bibinfo
  {pages} {e2220824120} (\bibinfo {year} {2023})}\BibitemShut {NoStop}%
\bibitem [{\citenamefont {Gabriel}\ and\ \citenamefont
  {Richert}(2023)}]{gabriel2023_comparing}%
  \BibitemOpen
  \bibfield  {author} {\bibinfo {author} {\bibfnamefont {J.~P.}\ \bibnamefont
  {Gabriel}}\ and\ \bibinfo {author} {\bibfnamefont {R.}~\bibnamefont
  {Richert}},\ }\bibfield  {title} {\enquote {\bibinfo {title} {{Comparing two
  sources of physical aging: Temperature vs electric field}},}\ }\href
  {https://doi.org/10.1063/5.0176957} {\bibfield  {journal} {\bibinfo
  {journal} {The Journal of Chemical Physics}\ }\textbf {\bibinfo {volume}
  {159}},\ \bibinfo {pages} {164502} (\bibinfo {year} {2023})}\BibitemShut
  {NoStop}%
\bibitem [{\citenamefont {Richert}\ and\ \citenamefont
  {Gabriel}(2023)}]{richert2023}%
  \BibitemOpen
  \bibfield  {author} {\bibinfo {author} {\bibfnamefont {R.}~\bibnamefont
  {Richert}}\ and\ \bibinfo {author} {\bibfnamefont {J.~P.}\ \bibnamefont
  {Gabriel}},\ }\bibfield  {title} {\enquote {\bibinfo {title} {{Fast vs slow
  physical aging of a glass forming liquid}},}\ }\href
  {https://doi.org/10.1063/5.0167766} {\bibfield  {journal} {\bibinfo
  {journal} {The Journal of Chemical Physics}\ }\textbf {\bibinfo {volume}
  {159}},\ \bibinfo {pages} {084504} (\bibinfo {year} {2023})}\BibitemShut
  {NoStop}%
\bibitem [{\citenamefont {Hénot}, \citenamefont {Nguyen},\ and\ \citenamefont
  {Ladieu}(2024)}]{henot2024}%
  \BibitemOpen
  \bibfield  {author} {\bibinfo {author} {\bibfnamefont {M.}~\bibnamefont
  {Hénot}}, \bibinfo {author} {\bibfnamefont {X.~A.}\ \bibnamefont {Nguyen}},\
  and\ \bibinfo {author} {\bibfnamefont {F.}~\bibnamefont {Ladieu}},\
  }\bibfield  {title} {\enquote {\bibinfo {title} {Crossing the frontier of
  validity of the material time approach in the aging of a molecular glass},}\
  }\href {https://doi.org/10.1021/acs.jpclett.4c00527} {\bibfield  {journal}
  {\bibinfo  {journal} {J. Phys. Chem. Lett.}\ }\textbf {\bibinfo {volume}
  {15}},\ \bibinfo {pages} {3170--3177} (\bibinfo {year} {2024})},\ \bibinfo
  {note} {pMID: 38478899}\BibitemShut {NoStop}%
\bibitem [{\citenamefont {Lancelotti}, \citenamefont {Zanotto},\ and\
  \citenamefont {Sen}(2024)}]{lancelotti2024}%
  \BibitemOpen
  \bibfield  {author} {\bibinfo {author} {\bibfnamefont {R.~F.}\ \bibnamefont
  {Lancelotti}}, \bibinfo {author} {\bibfnamefont {E.~D.}\ \bibnamefont
  {Zanotto}},\ and\ \bibinfo {author} {\bibfnamefont {S.}~\bibnamefont {Sen}},\
  }\bibfield  {title} {\enquote {\bibinfo {title} {Kinetics of physical aging
  of a silicate glass following temperature up-and down-jumps},}\ }\href@noop
  {} {\bibfield  {journal} {\bibinfo  {journal} {J. Chem. Phys.}\ }\textbf
  {\bibinfo {volume} {160}},\ \bibinfo {pages} {034504} (\bibinfo {year}
  {2024})}\BibitemShut {NoStop}%
\bibitem [{\citenamefont {Moch}, \citenamefont {Gainaru},\ and\ \citenamefont
  {Böhmer}(2024)}]{Moch2024}%
  \BibitemOpen
  \bibfield  {author} {\bibinfo {author} {\bibfnamefont {K.}~\bibnamefont
  {Moch}}, \bibinfo {author} {\bibfnamefont {C.}~\bibnamefont {Gainaru}},\ and\
  \bibinfo {author} {\bibfnamefont {R.}~\bibnamefont {Böhmer}},\ }\bibfield
  {title} {\enquote {\bibinfo {title} {{Nonlinear susceptibilities and
  higher-order responses related to physical aging: Wiener–Volterra approach
  and extended Tool–Narayanaswamy–Moynihan models}},}\ }\href
  {https://doi.org/10.1063/5.0207122} {\bibfield  {journal} {\bibinfo
  {journal} {The Journal of Chemical Physics}\ }\textbf {\bibinfo {volume}
  {161}},\ \bibinfo {pages} {014502} (\bibinfo {year} {2024})}\BibitemShut
  {NoStop}%
\bibitem [{\citenamefont {B{\"o}hmer}\ \emph {et~al.}(2024)\citenamefont
  {B{\"o}hmer}, \citenamefont {Gabriel}, \citenamefont {Costigliola},
  \citenamefont {Kociok}, \citenamefont {Hecksher}, \citenamefont {Dyre},\ and\
  \citenamefont {Blochowicz}}]{Bohmer2024}%
  \BibitemOpen
  \bibfield  {author} {\bibinfo {author} {\bibfnamefont {T.}~\bibnamefont
  {B{\"o}hmer}}, \bibinfo {author} {\bibfnamefont {J.~P.}\ \bibnamefont
  {Gabriel}}, \bibinfo {author} {\bibfnamefont {L.}~\bibnamefont
  {Costigliola}}, \bibinfo {author} {\bibfnamefont {J.-N.}\ \bibnamefont
  {Kociok}}, \bibinfo {author} {\bibfnamefont {T.}~\bibnamefont {Hecksher}},
  \bibinfo {author} {\bibfnamefont {J.~C.}\ \bibnamefont {Dyre}},\ and\
  \bibinfo {author} {\bibfnamefont {T.}~\bibnamefont {Blochowicz}},\ }\bibfield
   {title} {\enquote {\bibinfo {title} {Time reversibility during the ageing of
  materials},}\ }\href {https://doi.org/10.1038/s41567-023-02366-z} {\bibfield
  {journal} {\bibinfo  {journal} {Nature Physics}\ }\textbf {\bibinfo {volume}
  {20}},\ \bibinfo {pages} {637--645} (\bibinfo {year} {2024})}\BibitemShut
  {NoStop}%
\bibitem [{\citenamefont {Riechers}\ \emph {et~al.}(2024)\citenamefont
  {Riechers}, \citenamefont {Das}, \citenamefont {Dufresne}, \citenamefont
  {Derlet},\ and\ \citenamefont {Maass}}]{riechers2024}%
  \BibitemOpen
  \bibfield  {author} {\bibinfo {author} {\bibfnamefont {B.}~\bibnamefont
  {Riechers}}, \bibinfo {author} {\bibfnamefont {A.}~\bibnamefont {Das}},
  \bibinfo {author} {\bibfnamefont {E.}~\bibnamefont {Dufresne}}, \bibinfo
  {author} {\bibfnamefont {P.~M.}\ \bibnamefont {Derlet}},\ and\ \bibinfo
  {author} {\bibfnamefont {R.}~\bibnamefont {Maass}},\ }\bibfield  {title}
  {{\selectlanguage {English}\enquote {\bibinfo {title} {Intermittent cluster
  dynamics and temporal fractional diffusion in a bulk metallic glass},}\
  }}\href
  {https://www-proquest-com.ep.fjernadgang.kb.dk/scholarly-journals/intermittent-cluster-dynamics-temporal-fractional/docview/3087617785/se-2}
  {\bibfield  {journal} {\bibinfo  {journal} {Nature Communications}\ }\textbf
  {\bibinfo {volume} {15}},\ \bibinfo {pages} {6595} (\bibinfo {year}
  {2024})},\ \bibinfo {note} {copyright - © The Author(s) 2024. This work is
  published under http://creativecommons.org/licenses/by/4.0/ (the
  “License”). Notwithstanding the ProQuest Terms and Conditions, you may
  use this content in accordance with the terms of the License; Last updated -
  2024-08-04}\BibitemShut {NoStop}%
\bibitem [{\citenamefont {Macedo}\ and\ \citenamefont
  {Napolitano}(1967)}]{Macedo1967}%
  \BibitemOpen
  \bibfield  {author} {\bibinfo {author} {\bibfnamefont {P.~B.}\ \bibnamefont
  {Macedo}}\ and\ \bibinfo {author} {\bibfnamefont {A.}~\bibnamefont
  {Napolitano}},\ }\bibfield  {title} {\enquote {\bibinfo {title} {Effects of a
  distribution of volume relaxation times in annealing of bsc glass},}\
  }\href@noop {} {\bibfield  {journal} {\bibinfo  {journal} {Journal of
  Research of the National Bureau of Standards Section A-Physics}\ }\textbf
  {\bibinfo {volume} {71A}},\ \bibinfo {pages} {231--238} (\bibinfo {year}
  {1967})}\BibitemShut {NoStop}%
\bibitem [{\citenamefont {Huang}\ and\ \citenamefont {Paul}(2006)}]{Huang2006}%
  \BibitemOpen
  \bibfield  {author} {\bibinfo {author} {\bibfnamefont {Y.}~\bibnamefont
  {Huang}}\ and\ \bibinfo {author} {\bibfnamefont {D.}~\bibnamefont {Paul}},\
  }\bibfield  {title} {\enquote {\bibinfo {title} {Physical aging of thin
  glassy polymer films monitored by optical properties},}\ }\href
  {https://doi.org/10.1021/ma050533y} {\bibfield  {journal} {\bibinfo
  {journal} {Macromolecules}\ }\textbf {\bibinfo {volume} {39}},\ \bibinfo
  {pages} {1554--1559} (\bibinfo {year} {2006})}\BibitemShut {NoStop}%
\bibitem [{\citenamefont {Santore}, \citenamefont {Duran},\ and\ \citenamefont
  {McKenna}(1991)}]{Santore1991}%
  \BibitemOpen
  \bibfield  {author} {\bibinfo {author} {\bibfnamefont {M.~M.}\ \bibnamefont
  {Santore}}, \bibinfo {author} {\bibfnamefont {R.~S.}\ \bibnamefont {Duran}},\
  and\ \bibinfo {author} {\bibfnamefont {G.~B.}\ \bibnamefont {McKenna}},\
  }\bibfield  {title} {\enquote {\bibinfo {title} {Volume recovery in epoxy
  glasses subjected to torsional deformations: the question of rejuvenation},}\
  }\href {https://doi.org/https://doi.org/10.1016/0032-3861(91)90077-V}
  {\bibfield  {journal} {\bibinfo  {journal} {Polymer}\ }\textbf {\bibinfo
  {volume} {32}},\ \bibinfo {pages} {2377--2381} (\bibinfo {year}
  {1991})}\BibitemShut {NoStop}%
\bibitem [{\citenamefont {Struik}(1997)}]{Struik1997}%
  \BibitemOpen
  \bibfield  {author} {\bibinfo {author} {\bibfnamefont {L.~C.~E.}\
  \bibnamefont {Struik}},\ }\bibfield  {title} {\enquote {\bibinfo {title}
  {Volume-recovery theory: 1. {K}ovacs' $\tau$-effective paradox},}\
  }\href@noop {} {\bibfield  {journal} {\bibinfo  {journal} {Polymer}\ }\textbf
  {\bibinfo {volume} {38}},\ \bibinfo {pages} {4677 -- 4685} (\bibinfo {year}
  {1997})}\BibitemShut {NoStop}%
\bibitem [{\citenamefont {Cangialosi}\ \emph
  {et~al.}(2013{\natexlab{a}})\citenamefont {Cangialosi}, \citenamefont
  {Boucher}, \citenamefont {Alegría},\ and\ \citenamefont
  {Colmenero}}]{Cangialosi2013}%
  \BibitemOpen
  \bibfield  {author} {\bibinfo {author} {\bibfnamefont {D.}~\bibnamefont
  {Cangialosi}}, \bibinfo {author} {\bibfnamefont {V.~M.}\ \bibnamefont
  {Boucher}}, \bibinfo {author} {\bibfnamefont {A.}~\bibnamefont {Alegría}},\
  and\ \bibinfo {author} {\bibfnamefont {J.}~\bibnamefont {Colmenero}},\
  }\bibfield  {title} {\enquote {\bibinfo {title} {Volume recovery of
  polystyrene/silica nanocomposites},}\ }\href
  {https://doi.org/https://doi-org.ep.fjernadgang.kb.dk/10.1002/polb.23282}
  {\bibfield  {journal} {\bibinfo  {journal} {Journal of Polymer Science Part
  B: Polymer Physics}\ }\textbf {\bibinfo {volume} {51}},\ \bibinfo {pages}
  {847--853} (\bibinfo {year} {2013}{\natexlab{a}})}\BibitemShut {NoStop}%
\bibitem [{\citenamefont {Moynihan}\ \emph
  {et~al.}(1976{\natexlab{a}})\citenamefont {Moynihan}, \citenamefont {Macedo},
  \citenamefont {Montrose}, \citenamefont {Montrose}, \citenamefont {Gupta},
  \citenamefont {DeBolt}, \citenamefont {Dill}, \citenamefont {Dom},
  \citenamefont {Drake},\ and\ \citenamefont {Easteal}}]{Moynihan1976}%
  \BibitemOpen
  \bibfield  {author} {\bibinfo {author} {\bibfnamefont {C.}~\bibnamefont
  {Moynihan}}, \bibinfo {author} {\bibfnamefont {P.}~\bibnamefont {Macedo}},
  \bibinfo {author} {\bibfnamefont {C.}~\bibnamefont {Montrose}}, \bibinfo
  {author} {\bibfnamefont {C.}~\bibnamefont {Montrose}}, \bibinfo {author}
  {\bibfnamefont {P.}~\bibnamefont {Gupta}}, \bibinfo {author} {\bibfnamefont
  {M.}~\bibnamefont {DeBolt}}, \bibinfo {author} {\bibfnamefont
  {J.}~\bibnamefont {Dill}}, \bibinfo {author} {\bibfnamefont {B.}~\bibnamefont
  {Dom}}, \bibinfo {author} {\bibfnamefont {P.}~\bibnamefont {Drake}},\ and\
  \bibinfo {author} {\bibfnamefont {A.}~\bibnamefont {Easteal}},\ }\bibfield
  {title} {\enquote {\bibinfo {title} {Structural relaxation in vitreous
  materials},}\ }\href@noop {} {\bibfield  {journal} {\bibinfo  {journal} {Ann.
  N. Y. Acad. Sci.}\ }\textbf {\bibinfo {volume} {279}},\ \bibinfo {pages} {15}
  (\bibinfo {year} {1976}{\natexlab{a}})}\BibitemShut {NoStop}%
\bibitem [{\citenamefont {Kovacs}\ \emph {et~al.}(1979)\citenamefont {Kovacs},
  \citenamefont {Aklonis}, \citenamefont {Hutchinson},\ and\ \citenamefont
  {Ramos}}]{Kovacs1979}%
  \BibitemOpen
  \bibfield  {author} {\bibinfo {author} {\bibfnamefont {A.~J.}\ \bibnamefont
  {Kovacs}}, \bibinfo {author} {\bibfnamefont {J.~J.}\ \bibnamefont {Aklonis}},
  \bibinfo {author} {\bibfnamefont {J.~M.}\ \bibnamefont {Hutchinson}},\ and\
  \bibinfo {author} {\bibfnamefont {A.~R.}\ \bibnamefont {Ramos}},\ }\bibfield
  {title} {\enquote {\bibinfo {title} {{Isobaric Volume and Enthalpy Recovery
  of Glasses. II. A Transparent Multiparameter Theory}},}\ }\href@noop {}
  {\bibfield  {journal} {\bibinfo  {journal} {Journal of Polymer Science:
  Polymer Physics Edition}\ }\textbf {\bibinfo {volume} {17}},\ \bibinfo
  {pages} {1097--1162} (\bibinfo {year} {1979})}\BibitemShut {NoStop}%
\bibitem [{\citenamefont {Kovacs}, \citenamefont {Stratton},\ and\
  \citenamefont {Ferry}(1963)}]{Kovacs1963_dynamic}%
  \BibitemOpen
  \bibfield  {author} {\bibinfo {author} {\bibfnamefont {A.~J.}\ \bibnamefont
  {Kovacs}}, \bibinfo {author} {\bibfnamefont {R.~A.}\ \bibnamefont
  {Stratton}},\ and\ \bibinfo {author} {\bibfnamefont {J.~D.}\ \bibnamefont
  {Ferry}},\ }\bibfield  {title} {\enquote {\bibinfo {title} {Dynamic
  mechanical properties of polyvinyl acetate in shear in the glass transition
  temperature range},}\ }\href@noop {} {\bibfield  {journal} {\bibinfo
  {journal} {J. Phys. Chem.}\ }\textbf {\bibinfo {volume} {67}},\ \bibinfo
  {pages} {152--161} (\bibinfo {year} {1963})}\BibitemShut {NoStop}%
\bibitem [{\citenamefont {Echeverria}\ \emph {et~al.}(1995)\citenamefont
  {Echeverria}, \citenamefont {Su}, \citenamefont {Simon},\ and\ \citenamefont
  {Plazek}}]{Echeverria1995}%
  \BibitemOpen
  \bibfield  {author} {\bibinfo {author} {\bibfnamefont {I.}~\bibnamefont
  {Echeverria}}, \bibinfo {author} {\bibfnamefont {P.-C.}\ \bibnamefont {Su}},
  \bibinfo {author} {\bibfnamefont {S.~L.}\ \bibnamefont {Simon}},\ and\
  \bibinfo {author} {\bibfnamefont {D.~J.}\ \bibnamefont {Plazek}},\ }\bibfield
   {title} {\enquote {\bibinfo {title} {{Physical aging of a polyetherimide:
  Creep and DSC measurements}},}\ }\href
  {https://doi.org/https://doi-org.ep.fjernadgang.kb.dk/10.1002/polb.1995.090331717}
  {\bibfield  {journal} {\bibinfo  {journal} {Journal of Polymer Science Part
  B: Polymer Physics}\ }\textbf {\bibinfo {volume} {33}},\ \bibinfo {pages}
  {2457--2468} (\bibinfo {year} {1995})}\BibitemShut {NoStop}%
\bibitem [{\citenamefont {Soloukhin}\ \emph {et~al.}(2003)\citenamefont
  {Soloukhin}, \citenamefont {Brokken-Zijp}, \citenamefont {van Asselen},\ and\
  \citenamefont {de~With}}]{Soloukhin2003}%
  \BibitemOpen
  \bibfield  {author} {\bibinfo {author} {\bibfnamefont {V.~A.}\ \bibnamefont
  {Soloukhin}}, \bibinfo {author} {\bibfnamefont {J.~C.~M.}\ \bibnamefont
  {Brokken-Zijp}}, \bibinfo {author} {\bibfnamefont {O.~L.~J.}\ \bibnamefont
  {van Asselen}},\ and\ \bibinfo {author} {\bibfnamefont {G.}~\bibnamefont
  {de~With}},\ }\bibfield  {title} {\enquote {\bibinfo {title} {Physical aging
  of polycarbonate: Elastic modulus, hardness, creep, endothermic peak,
  molecular weight distribution, and infrared data},}\ }\href
  {https://doi.org/10.1021/ma0342980} {\bibfield  {journal} {\bibinfo
  {journal} {Macromolecules}\ }\textbf {\bibinfo {volume} {36}},\ \bibinfo
  {pages} {7585--7597} (\bibinfo {year} {2003})}\BibitemShut {NoStop}%
\bibitem [{\citenamefont {Huang}\ and\ \citenamefont {Paul}(2004)}]{Huang2004}%
  \BibitemOpen
  \bibfield  {author} {\bibinfo {author} {\bibfnamefont {Y.}~\bibnamefont
  {Huang}}\ and\ \bibinfo {author} {\bibfnamefont {D.~R.}\ \bibnamefont
  {Paul}},\ }\bibfield  {title} {\enquote {\bibinfo {title} {Physical aging of
  thin glassy polymer films monitored by gas permeability},}\ }\href
  {https://doi.org/https://doi.org/10.1016/j.polymer.2004.10.019} {\bibfield
  {journal} {\bibinfo  {journal} {Polymer}\ }\textbf {\bibinfo {volume} {45}},\
  \bibinfo {pages} {8377--8393} (\bibinfo {year} {2004})}\BibitemShut {NoStop}%
\bibitem [{\citenamefont {Leheny}\ and\ \citenamefont
  {Nagel}(1998)}]{leheny1998}%
  \BibitemOpen
  \bibfield  {author} {\bibinfo {author} {\bibfnamefont {R.~L.}\ \bibnamefont
  {Leheny}}\ and\ \bibinfo {author} {\bibfnamefont {S.~R.}\ \bibnamefont
  {Nagel}},\ }\bibfield  {title} {\enquote {\bibinfo {title} {Frequency-domain
  study of physical aging in a simple liquid},}\ }\href@noop {} {\bibfield
  {journal} {\bibinfo  {journal} {Phys. Rev. B}\ }\textbf {\bibinfo {volume}
  {57}},\ \bibinfo {pages} {5154} (\bibinfo {year} {1998})}\BibitemShut
  {NoStop}%
\bibitem [{\citenamefont {Lunkenheimer}\ \emph {et~al.}(2005)\citenamefont
  {Lunkenheimer}, \citenamefont {Wehn}, \citenamefont {Schneider},\ and\
  \citenamefont {Loidl}}]{lunkenheimer2005}%
  \BibitemOpen
  \bibfield  {author} {\bibinfo {author} {\bibfnamefont {P.}~\bibnamefont
  {Lunkenheimer}}, \bibinfo {author} {\bibfnamefont {R.}~\bibnamefont {Wehn}},
  \bibinfo {author} {\bibfnamefont {U.}~\bibnamefont {Schneider}},\ and\
  \bibinfo {author} {\bibfnamefont {A.}~\bibnamefont {Loidl}},\ }\bibfield
  {title} {\enquote {\bibinfo {title} {Glassy aging dynamics},}\ }\href@noop {}
  {\bibfield  {journal} {\bibinfo  {journal} {Phys. Rev. Lett.}\ }\textbf
  {\bibinfo {volume} {95}},\ \bibinfo {pages} {055702} (\bibinfo {year}
  {2005})}\BibitemShut {NoStop}%
\bibitem [{\citenamefont {Richert}\ \emph {et~al.}(2013)\citenamefont
  {Richert}, \citenamefont {Lunkenheimer}, \citenamefont {Kastner},\ and\
  \citenamefont {Loidl}}]{Richert2013}%
  \BibitemOpen
  \bibfield  {author} {\bibinfo {author} {\bibfnamefont {R.}~\bibnamefont
  {Richert}}, \bibinfo {author} {\bibfnamefont {P.}~\bibnamefont
  {Lunkenheimer}}, \bibinfo {author} {\bibfnamefont {S.}~\bibnamefont
  {Kastner}},\ and\ \bibinfo {author} {\bibfnamefont {A.}~\bibnamefont
  {Loidl}},\ }\bibfield  {title} {\enquote {\bibinfo {title} {On the derivation
  of equilibrium relaxation times from aging experiments},}\ }\href
  {https://doi.org/dx.doi.org/10.1021/jp311149n} {\bibfield  {journal}
  {\bibinfo  {journal} {The Journal of Physical Chemistry B}\ }\textbf
  {\bibinfo {volume} {117}},\ \bibinfo {pages} {12689--12694} (\bibinfo {year}
  {2013})}\BibitemShut {NoStop}%
\bibitem [{\citenamefont {Dambon}\ \emph {et~al.}(2009)\citenamefont {Dambon},
  \citenamefont {Wang}, \citenamefont {Klocke}, \citenamefont {Pongs},
  \citenamefont {Bresseler}, \citenamefont {Chen},\ and\ \citenamefont
  {Yi}}]{Dambon2009}%
  \BibitemOpen
  \bibfield  {author} {\bibinfo {author} {\bibfnamefont {O.}~\bibnamefont
  {Dambon}}, \bibinfo {author} {\bibfnamefont {F.}~\bibnamefont {Wang}},
  \bibinfo {author} {\bibfnamefont {F.}~\bibnamefont {Klocke}}, \bibinfo
  {author} {\bibfnamefont {G.}~\bibnamefont {Pongs}}, \bibinfo {author}
  {\bibfnamefont {B.}~\bibnamefont {Bresseler}}, \bibinfo {author}
  {\bibfnamefont {Y.}~\bibnamefont {Chen}},\ and\ \bibinfo {author}
  {\bibfnamefont {A.~Y.}\ \bibnamefont {Yi}},\ }\bibfield  {title} {\enquote
  {\bibinfo {title} {{Efficient mold manufacturing for precision glass
  molding}},}\ }\href {https://doi.org/10.1116/1.3056171} {\bibfield  {journal}
  {\bibinfo  {journal} {Journal of Vacuum Science \& Technology B:
  Microelectronics and Nanometer Structures Processing, Measurement, and
  Phenomena}\ }\textbf {\bibinfo {volume} {27}},\ \bibinfo {pages} {1445--1449}
  (\bibinfo {year} {2009})}\BibitemShut {NoStop}%
\bibitem [{\citenamefont {Ming}\ \emph {et~al.}(2020)\citenamefont {Ming},
  \citenamefont {Chen}, \citenamefont {Du}, \citenamefont {Zhang},
  \citenamefont {Zhang}, \citenamefont {He}, \citenamefont {Ma},\ and\
  \citenamefont {Shen}}]{Ming2020}%
  \BibitemOpen
  \bibfield  {author} {\bibinfo {author} {\bibfnamefont {W.}~\bibnamefont
  {Ming}}, \bibinfo {author} {\bibfnamefont {Z.}~\bibnamefont {Chen}}, \bibinfo
  {author} {\bibfnamefont {J.}~\bibnamefont {Du}}, \bibinfo {author}
  {\bibfnamefont {Z.}~\bibnamefont {Zhang}}, \bibinfo {author} {\bibfnamefont
  {G.}~\bibnamefont {Zhang}}, \bibinfo {author} {\bibfnamefont
  {W.}~\bibnamefont {He}}, \bibinfo {author} {\bibfnamefont {J.}~\bibnamefont
  {Ma}},\ and\ \bibinfo {author} {\bibfnamefont {F.}~\bibnamefont {Shen}},\
  }\bibfield  {title} {\enquote {\bibinfo {title} {A comprehensive review of
  theory and technology of glass molding process},}\ }\href@noop {} {\bibfield
  {journal} {\bibinfo  {journal} {Internat. J. Adv. Manufact. Tech.}\ }\textbf
  {\bibinfo {volume} {107}},\ \bibinfo {pages} {2671–2706} (\bibinfo {year}
  {2020})}\BibitemShut {NoStop}%
\bibitem [{\citenamefont {Jiang}\ \emph {et~al.}(2022)\citenamefont {Jiang},
  \citenamefont {Tovar}, \citenamefont {Staasmeyer}, \citenamefont
  {Friedrichs}, \citenamefont {Grunwald},\ and\ \citenamefont
  {Bergs}}]{Jiang2022}%
  \BibitemOpen
  \bibfield  {author} {\bibinfo {author} {\bibfnamefont {C.}~\bibnamefont
  {Jiang}}, \bibinfo {author} {\bibfnamefont {C.~M.}\ \bibnamefont {Tovar}},
  \bibinfo {author} {\bibfnamefont {J.-H.}\ \bibnamefont {Staasmeyer}},
  \bibinfo {author} {\bibfnamefont {M.}~\bibnamefont {Friedrichs}}, \bibinfo
  {author} {\bibfnamefont {T.}~\bibnamefont {Grunwald}},\ and\ \bibinfo
  {author} {\bibfnamefont {T.}~\bibnamefont {Bergs}},\ }\bibfield  {title}
  {\enquote {\bibinfo {title} {{Simulation of the Refractive Index Variation
  and Validation of the Form Deviation in Precisely Molded Chalcogenide Glass
  Lenses (IRG 26) Considering the Stress and Structure Relaxation}},}\ }\href
  {https://doi.org/10.3390/ma15196756} {\bibfield  {journal} {\bibinfo
  {journal} {Materials}\ }\textbf {\bibinfo {volume} {15}} (\bibinfo {year}
  {2022}),\ 10.3390/ma15196756}\BibitemShut {NoStop}%
\bibitem [{\citenamefont {Vu}\ \emph {et~al.}(2022)\citenamefont {Vu},
  \citenamefont {Avila~Hernandez}, \citenamefont {Grunwald},\ and\
  \citenamefont {Bergs}}]{Vu2022}%
  \BibitemOpen
  \bibfield  {author} {\bibinfo {author} {\bibfnamefont {A.~T.}\ \bibnamefont
  {Vu}}, \bibinfo {author} {\bibfnamefont {R.~d. l.~A.}\ \bibnamefont
  {Avila~Hernandez}}, \bibinfo {author} {\bibfnamefont {T.}~\bibnamefont
  {Grunwald}},\ and\ \bibinfo {author} {\bibfnamefont {T.}~\bibnamefont
  {Bergs}},\ }\bibfield  {title} {\enquote {\bibinfo {title} {Modeling
  nonequilibrium thermoviscoelastic material behaviors of glass in
  nonisothermal glass molding},}\ }\href@noop {} {\bibfield  {journal}
  {\bibinfo  {journal} {J. Am. Ceram. Soc.}\ }\textbf {\bibinfo {volume}
  {105}},\ \bibinfo {pages} {6799--6815} (\bibinfo {year} {2022})}\BibitemShut
  {NoStop}%
\bibitem [{\citenamefont {Jha}\ \emph {et~al.}(2012)\citenamefont {Jha},
  \citenamefont {Richards}, \citenamefont {Jose}, \citenamefont {Fernandez},
  \citenamefont {Hill}, \citenamefont {Lousteau},\ and\ \citenamefont
  {Joshi}}]{Jha2012}%
  \BibitemOpen
  \bibfield  {author} {\bibinfo {author} {\bibfnamefont {A.}~\bibnamefont
  {Jha}}, \bibinfo {author} {\bibfnamefont {B.~D.~O.}\ \bibnamefont
  {Richards}}, \bibinfo {author} {\bibfnamefont {G.}~\bibnamefont {Jose}},
  \bibinfo {author} {\bibfnamefont {T.~T.}\ \bibnamefont {Fernandez}}, \bibinfo
  {author} {\bibfnamefont {C.~J.}\ \bibnamefont {Hill}}, \bibinfo {author}
  {\bibfnamefont {J.}~\bibnamefont {Lousteau}},\ and\ \bibinfo {author}
  {\bibfnamefont {P.}~\bibnamefont {Joshi}},\ }\bibfield  {title} {\enquote
  {\bibinfo {title} {Review on structural, thermal, optical and spectroscopic
  properties of tellurium oxide based glasses for fibre optic and waveguide
  applications},}\ }\href {https://doi.org/10.1179/1743280412Y.0000000005}
  {\bibfield  {journal} {\bibinfo  {journal} {International Materials Reviews}\
  }\textbf {\bibinfo {volume} {57}},\ \bibinfo {pages} {357--382} (\bibinfo
  {year} {2012})}\BibitemShut {NoStop}%
\bibitem [{\citenamefont {Ellison}\ and\ \citenamefont
  {Cornejo}(2010)}]{Ellison2010}%
  \BibitemOpen
  \bibfield  {author} {\bibinfo {author} {\bibfnamefont {A.}~\bibnamefont
  {Ellison}}\ and\ \bibinfo {author} {\bibfnamefont {I.~A.}\ \bibnamefont
  {Cornejo}},\ }\bibfield  {title} {\enquote {\bibinfo {title} {Glass
  substrates for liquid crystal displays},}\ }\href
  {https://doi.org/https://doi-org.ep.fjernadgang.kb.dk/10.1111/j.2041-1294.2010.00009.x}
  {\bibfield  {journal} {\bibinfo  {journal} {International Journal of Applied
  Glass Science}\ }\textbf {\bibinfo {volume} {1}},\ \bibinfo {pages} {87--103}
  (\bibinfo {year} {2010})}\BibitemShut {NoStop}%
\bibitem [{\citenamefont {Weyhe}\ \emph {et~al.}(2023)\citenamefont {Weyhe},
  \citenamefont {Andersen}, \citenamefont {Mikkelsen},\ and\ \citenamefont
  {Yu}}]{Weyhe2023}%
  \BibitemOpen
  \bibfield  {author} {\bibinfo {author} {\bibfnamefont {A.~T.}\ \bibnamefont
  {Weyhe}}, \bibinfo {author} {\bibfnamefont {E.}~\bibnamefont {Andersen}},
  \bibinfo {author} {\bibfnamefont {R.}~\bibnamefont {Mikkelsen}},\ and\
  \bibinfo {author} {\bibfnamefont {D.}~\bibnamefont {Yu}},\ }\bibfield
  {title} {\enquote {\bibinfo {title} {Accelerated physical aging of four pet
  copolyesters: Enthalpy relaxation and yield behaviour},}\ }\href
  {https://doi.org/https://doi.org/10.1016/j.polymer.2023.125987} {\bibfield
  {journal} {\bibinfo  {journal} {Polymer}\ }\textbf {\bibinfo {volume}
  {278}},\ \bibinfo {pages} {125987} (\bibinfo {year} {2023})}\BibitemShut
  {NoStop}%
\bibitem [{\citenamefont {Merrick}, \citenamefont {Sujanani},\ and\
  \citenamefont {Freeman}(2020)}]{Merrick2020}%
  \BibitemOpen
  \bibfield  {author} {\bibinfo {author} {\bibfnamefont {M.~M.}\ \bibnamefont
  {Merrick}}, \bibinfo {author} {\bibfnamefont {R.}~\bibnamefont {Sujanani}},\
  and\ \bibinfo {author} {\bibfnamefont {B.~D.}\ \bibnamefont {Freeman}},\
  }\bibfield  {title} {\enquote {\bibinfo {title} {Glassy polymers: Historical
  findings, membrane applications, and unresolved questions regarding physical
  aging},}\ }\href
  {https://doi.org/https://doi.org/10.1016/j.polymer.2020.123176} {\bibfield
  {journal} {\bibinfo  {journal} {Polymer}\ }\textbf {\bibinfo {volume}
  {211}},\ \bibinfo {pages} {123176} (\bibinfo {year} {2020})}\BibitemShut
  {NoStop}%
\bibitem [{\citenamefont {Boccaccini}\ \emph {et~al.}(2010)\citenamefont
  {Boccaccini}, \citenamefont {Erol}, \citenamefont {Stark}, \citenamefont
  {Mohn}, \citenamefont {Hong},\ and\ \citenamefont {Mano}}]{Boccaccini2010}%
  \BibitemOpen
  \bibfield  {author} {\bibinfo {author} {\bibfnamefont {A.~R.}\ \bibnamefont
  {Boccaccini}}, \bibinfo {author} {\bibfnamefont {M.}~\bibnamefont {Erol}},
  \bibinfo {author} {\bibfnamefont {W.~J.}\ \bibnamefont {Stark}}, \bibinfo
  {author} {\bibfnamefont {D.}~\bibnamefont {Mohn}}, \bibinfo {author}
  {\bibfnamefont {Z.}~\bibnamefont {Hong}},\ and\ \bibinfo {author}
  {\bibfnamefont {J.~F.}\ \bibnamefont {Mano}},\ }\bibfield  {title} {\enquote
  {\bibinfo {title} {Polymer/bioactive glass nanocomposites for biomedical
  applications: A review},}\ }\href
  {https://doi.org/https://doi.org/10.1016/j.compscitech.2010.06.002}
  {\bibfield  {journal} {\bibinfo  {journal} {Composites Science and
  Technology}\ }\textbf {\bibinfo {volume} {70}},\ \bibinfo {pages}
  {1764--1776} (\bibinfo {year} {2010})},\ \bibinfo {note} {iCCM-17: Composites
  In Biomedical Applications}\BibitemShut {NoStop}%
\bibitem [{\citenamefont {Tool}(1946)}]{tool1946}%
  \BibitemOpen
  \bibfield  {author} {\bibinfo {author} {\bibfnamefont {A.~Q.}\ \bibnamefont
  {Tool}},\ }\bibfield  {title} {\enquote {\bibinfo {title} {Relation between
  inelastic deformability and thermal expansion of glass in its annealing
  range},}\ }\href@noop {} {\bibfield  {journal} {\bibinfo  {journal} {J. Am.
  Ceram. Soc.}\ }\textbf {\bibinfo {volume} {29}},\ \bibinfo {pages} {240}
  (\bibinfo {year} {1946})}\BibitemShut {NoStop}%
\bibitem [{\citenamefont {Narayanaswamy}(1971)}]{narayanaswamy1971}%
  \BibitemOpen
  \bibfield  {author} {\bibinfo {author} {\bibfnamefont {O.}~\bibnamefont
  {Narayanaswamy}},\ }\bibfield  {title} {\enquote {\bibinfo {title} {A model
  of structural relaxation in glass},}\ }\href@noop {} {\bibfield  {journal}
  {\bibinfo  {journal} {J. Am. Ceram. Soc.}\ }\textbf {\bibinfo {volume}
  {54}},\ \bibinfo {pages} {491} (\bibinfo {year} {1971})}\BibitemShut
  {NoStop}%
\bibitem [{\citenamefont {Struik}(1977)}]{Struik1977}%
  \BibitemOpen
  \bibfield  {author} {\bibinfo {author} {\bibfnamefont {L.~C.~E.}\
  \bibnamefont {Struik}},\ }\bibfield  {title} {\enquote {\bibinfo {title}
  {Physical aging in plastics and other glassy materials},}\ }\href@noop {}
  {\bibfield  {journal} {\bibinfo  {journal} {Polym. Eng. Sci.}\ }\textbf
  {\bibinfo {volume} {17}},\ \bibinfo {pages} {165--173} (\bibinfo {year}
  {1977})}\BibitemShut {NoStop}%
\bibitem [{\citenamefont {Hutchinson}(1995)}]{Hutchinson1995}%
  \BibitemOpen
  \bibfield  {author} {\bibinfo {author} {\bibfnamefont {J.~M.}\ \bibnamefont
  {Hutchinson}},\ }\bibfield  {title} {\enquote {\bibinfo {title} {Physical
  aging of polymers},}\ }\href
  {https://doi.org/https://doi.org/10.1016/0079-6700(94)00001-I} {\bibfield
  {journal} {\bibinfo  {journal} {Progress in Polymer Science}\ }\textbf
  {\bibinfo {volume} {20}},\ \bibinfo {pages} {703--760} (\bibinfo {year}
  {1995})}\BibitemShut {NoStop}%
\bibitem [{\citenamefont {McKenna}(2020)}]{McKenna2020}%
  \BibitemOpen
  \bibfield  {author} {\bibinfo {author} {\bibfnamefont {G.~B.}\ \bibnamefont
  {McKenna}},\ }\bibfield  {title} {\enquote {\bibinfo {title} {Looking at the
  glass transition: Challenges of extreme time scales and other interesting
  problems},}\ }\href@noop {} {\bibfield  {journal} {\bibinfo  {journal}
  {Rubber Chem. Technol.}\ }\textbf {\bibinfo {volume} {93}},\ \bibinfo {pages}
  {79--120} (\bibinfo {year} {2020})}\BibitemShut {NoStop}%
\bibitem [{\citenamefont {Boucher}\ \emph {et~al.}(2011)\citenamefont
  {Boucher}, \citenamefont {Cangialosi}, \citenamefont {Alegría},\ and\
  \citenamefont {Colmenero‡†§}}]{Boucher2011}%
  \BibitemOpen
  \bibfield  {author} {\bibinfo {author} {\bibfnamefont {V.~M.}\ \bibnamefont
  {Boucher}}, \bibinfo {author} {\bibfnamefont {D.}~\bibnamefont {Cangialosi}},
  \bibinfo {author} {\bibfnamefont {A.}~\bibnamefont {Alegría}},\ and\
  \bibinfo {author} {\bibfnamefont {J.}~\bibnamefont {Colmenero‡†§}},\
  }\bibfield  {title} {\enquote {\bibinfo {title} {Enthalpy recovery of glassy
  polymers: Dramatic deviations from the extrapolated liquidlike behavior},}\
  }\href {https://doi.org/https://doi.org/10.1021/ma2018233} {\bibfield
  {journal} {\bibinfo  {journal} {Macromolecules}\ }\textbf {\bibinfo {volume}
  {44}},\ \bibinfo {pages} {8333--8342} (\bibinfo {year} {2011})}\BibitemShut
  {NoStop}%
\bibitem [{\citenamefont {Eulatea}\ and\ \citenamefont
  {Cangialosi}(2018)}]{Eulatea2018}%
  \BibitemOpen
  \bibfield  {author} {\bibinfo {author} {\bibfnamefont {N.~G. P.-D.}\
  \bibnamefont {Eulatea}}\ and\ \bibinfo {author} {\bibfnamefont
  {D.}~\bibnamefont {Cangialosi}},\ }\bibfield  {title} {\enquote {\bibinfo
  {title} {The very long-term physical aging of glassy polymers},}\ }\href
  {https://doi.org/DOI: 10.1039/c8cp01940a} {\bibfield  {journal} {\bibinfo
  {journal} {Phys. Chem. Chem. Phys.,}\ }\textbf {\bibinfo {volume} {20}},\
  \bibinfo {pages} {12356--12361} (\bibinfo {year} {2018})}\BibitemShut
  {NoStop}%
\bibitem [{\citenamefont {Cangialosi}\ \emph
  {et~al.}(2013{\natexlab{b}})\citenamefont {Cangialosi}, \citenamefont
  {Boucher}, \citenamefont {Alegr\'{\i}a},\ and\ \citenamefont
  {Colmenero}}]{Cangialosi2013b}%
  \BibitemOpen
  \bibfield  {author} {\bibinfo {author} {\bibfnamefont {D.}~\bibnamefont
  {Cangialosi}}, \bibinfo {author} {\bibfnamefont {V.~M.}\ \bibnamefont
  {Boucher}}, \bibinfo {author} {\bibfnamefont {A.}~\bibnamefont
  {Alegr\'{\i}a}},\ and\ \bibinfo {author} {\bibfnamefont {J.}~\bibnamefont
  {Colmenero}},\ }\bibfield  {title} {\enquote {\bibinfo {title} {Direct
  evidence of two equilibration mechanisms in glassy polymers},}\ }\href
  {https://doi.org/10.1103/PhysRevLett.111.095701} {\bibfield  {journal}
  {\bibinfo  {journal} {Phys. Rev. Lett.}\ }\textbf {\bibinfo {volume} {111}},\
  \bibinfo {pages} {095701} (\bibinfo {year} {2013}{\natexlab{b}})}\BibitemShut
  {NoStop}%
\bibitem [{\citenamefont {Mehri}, \citenamefont {Ingebrigtsen},\ and\
  \citenamefont {Dyre}(2021)}]{mehri2021}%
  \BibitemOpen
  \bibfield  {author} {\bibinfo {author} {\bibfnamefont {S.}~\bibnamefont
  {Mehri}}, \bibinfo {author} {\bibfnamefont {T.~S.}\ \bibnamefont
  {Ingebrigtsen}},\ and\ \bibinfo {author} {\bibfnamefont {J.~C.}\ \bibnamefont
  {Dyre}},\ }\bibfield  {title} {\enquote {\bibinfo {title} {{Single-parameter
  aging in a binary Lennard-Jones system}},}\ }\href
  {https://doi.org/10.1063/5.0039250} {\bibfield  {journal} {\bibinfo
  {journal} {The Journal of Chemical Physics}\ }\textbf {\bibinfo {volume}
  {154}},\ \bibinfo {pages} {094504} (\bibinfo {year} {2021})}\BibitemShut
  {NoStop}%
\bibitem [{\citenamefont {Alba-Simionesco}, \citenamefont {Kivelson},\ and\
  \citenamefont {Tarjus}(2002)}]{Alba2002}%
  \BibitemOpen
  \bibfield  {author} {\bibinfo {author} {\bibfnamefont {C.}~\bibnamefont
  {Alba-Simionesco}}, \bibinfo {author} {\bibfnamefont {D.}~\bibnamefont
  {Kivelson}},\ and\ \bibinfo {author} {\bibfnamefont {G.}~\bibnamefont
  {Tarjus}},\ }\bibfield  {title} {\enquote {\bibinfo {title} {Temperature,
  density, and pressure dependence of relaxation times in supercooled
  liquids},}\ }\href@noop {} {\bibfield  {journal} {\bibinfo  {journal} {J.
  Chem. Phys.}\ }\textbf {\bibinfo {volume} {116}},\ \bibinfo {pages} {5033}
  (\bibinfo {year} {2002})}\BibitemShut {NoStop}%
\bibitem [{\citenamefont {Tarjus}\ \emph {et~al.}(2004)\citenamefont {Tarjus},
  \citenamefont {Kivelson}, \citenamefont {Mossa},\ and\ \citenamefont
  {Alba-Simionesco}}]{Tarjus2004}%
  \BibitemOpen
  \bibfield  {author} {\bibinfo {author} {\bibfnamefont {G.}~\bibnamefont
  {Tarjus}}, \bibinfo {author} {\bibfnamefont {D.}~\bibnamefont {Kivelson}},
  \bibinfo {author} {\bibfnamefont {S.}~\bibnamefont {Mossa}},\ and\ \bibinfo
  {author} {\bibfnamefont {C.}~\bibnamefont {Alba-Simionesco}},\ }\bibfield
  {title} {\enquote {\bibinfo {title} {Disentangling density and temperature
  effects in the viscous slowing down of glassforming liquids},}\ }\href@noop
  {} {\bibfield  {journal} {\bibinfo  {journal} {J. Chem. Phys.}\ }\textbf
  {\bibinfo {volume} {120}},\ \bibinfo {pages} {6135--6141} (\bibinfo {year}
  {2004})}\BibitemShut {NoStop}%
\bibitem [{\citenamefont {Casalini}\ and\ \citenamefont
  {Roland}(2004)}]{Casalini2004}%
  \BibitemOpen
  \bibfield  {author} {\bibinfo {author} {\bibfnamefont {R.}~\bibnamefont
  {Casalini}}\ and\ \bibinfo {author} {\bibfnamefont {C.~M.}\ \bibnamefont
  {Roland}},\ }\bibfield  {title} {\enquote {\bibinfo {title} {Thermodynamical
  scaling of the glass transition dynamics},}\ }\href@noop {} {\bibfield
  {journal} {\bibinfo  {journal} {Phys. Rev. E}\ }\textbf {\bibinfo {volume}
  {69}},\ \bibinfo {pages} {062501} (\bibinfo {year} {2004})}\BibitemShut
  {NoStop}%
\bibitem [{\citenamefont {Roland}\ \emph {et~al.}(2005)\citenamefont {Roland},
  \citenamefont {Hensel-Bielowka}, \citenamefont {Paluch},\ and\ \citenamefont
  {Casalini}}]{Roland2005}%
  \BibitemOpen
  \bibfield  {author} {\bibinfo {author} {\bibfnamefont {C.~M.}\ \bibnamefont
  {Roland}}, \bibinfo {author} {\bibfnamefont {S.}~\bibnamefont
  {Hensel-Bielowka}}, \bibinfo {author} {\bibfnamefont {M.}~\bibnamefont
  {Paluch}},\ and\ \bibinfo {author} {\bibfnamefont {R.}~\bibnamefont
  {Casalini}},\ }\bibfield  {title} {\enquote {\bibinfo {title} {Supercooled
  dynamics of glass-forming liquids and polymers under hydrostatic pressure},}\
  }\href@noop {} {\bibfield  {journal} {\bibinfo  {journal} {Rep. Prog. Phys.}\
  }\textbf {\bibinfo {volume} {68}},\ \bibinfo {pages} {1405--1478} (\bibinfo
  {year} {2005})}\BibitemShut {NoStop}%
\bibitem [{\citenamefont {Niss}(2017)}]{niss2017}%
  \BibitemOpen
  \bibfield  {author} {\bibinfo {author} {\bibfnamefont {K.}~\bibnamefont
  {Niss}},\ }\bibfield  {title} {\enquote {\bibinfo {title} {Mapping isobaric
  aging onto the equilibrium phase diagram},}\ }\href@noop {} {\bibfield
  {journal} {\bibinfo  {journal} {Phys. Rev. Lett.}\ }\textbf {\bibinfo
  {volume} {119}},\ \bibinfo {pages} {115703} (\bibinfo {year}
  {2017})}\BibitemShut {NoStop}%
\bibitem [{\citenamefont {Niss}(2022)}]{niss2022}%
  \BibitemOpen
  \bibfield  {author} {\bibinfo {author} {\bibfnamefont {K.}~\bibnamefont
  {Niss}},\ }\bibfield  {title} {\enquote {\bibinfo {title} {A density scaling
  conjecture for aging glasses},}\ }\href@noop {} {\bibfield  {journal}
  {\bibinfo  {journal} {J. Chem. Phys.}\ }\textbf {\bibinfo {volume} {157}},\
  \bibinfo {pages} {054503} (\bibinfo {year} {2022})}\BibitemShut {NoStop}%
\bibitem [{\citenamefont {Hecksher}, \citenamefont {Olsen},\ and\ \citenamefont
  {Dyre}(2015)}]{hecksher2015}%
  \BibitemOpen
  \bibfield  {author} {\bibinfo {author} {\bibfnamefont {T.}~\bibnamefont
  {Hecksher}}, \bibinfo {author} {\bibfnamefont {N.~B.}\ \bibnamefont
  {Olsen}},\ and\ \bibinfo {author} {\bibfnamefont {J.~C.}\ \bibnamefont
  {Dyre}},\ }\bibfield  {title} {\enquote {\bibinfo {title} {Communication:
  Direct tests of single-parameter aging},}\ }\href@noop {} {\bibfield
  {journal} {\bibinfo  {journal} {J. Chem. Phys.}\ }\textbf {\bibinfo {volume}
  {142}},\ \bibinfo {pages} {241103} (\bibinfo {year} {2015})}\BibitemShut
  {NoStop}%
\bibitem [{\citenamefont {Niss}, \citenamefont {Dyre},\ and\ \citenamefont
  {Hecksher}(2020)}]{niss2020}%
  \BibitemOpen
  \bibfield  {author} {\bibinfo {author} {\bibfnamefont {K.}~\bibnamefont
  {Niss}}, \bibinfo {author} {\bibfnamefont {J.~C.}\ \bibnamefont {Dyre}},\
  and\ \bibinfo {author} {\bibfnamefont {T.}~\bibnamefont {Hecksher}},\
  }\bibfield  {title} {\enquote {\bibinfo {title} {{Long-time structural
  relaxation of glass-forming liquids: Simple or stretched exponential?}}}\
  }\href {https://doi.org/10.1063/1.5142189} {\bibfield  {journal} {\bibinfo
  {journal} {The Journal of Chemical Physics}\ }\textbf {\bibinfo {volume}
  {152}},\ \bibinfo {pages} {041103} (\bibinfo {year} {2020})}\BibitemShut
  {NoStop}%
\bibitem [{\citenamefont {Kovacs}(1963)}]{Kovacs1963_transition}%
  \BibitemOpen
  \bibfield  {author} {\bibinfo {author} {\bibfnamefont {A.~J.}\ \bibnamefont
  {Kovacs}},\ }\bibfield  {title} {\enquote {\bibinfo {title} {{Transition
  vitreuse dans les polymeres amorphes. Etude phenomenologique.}}}\ }\href@noop
  {} {\bibfield  {journal} {\bibinfo  {journal} {Fortschritte der
  Hochpolymeren-Forschung}\ }\textbf {\bibinfo {volume} {3}},\ \bibinfo {pages}
  {394--507} (\bibinfo {year} {1963})}\BibitemShut {NoStop}%
\bibitem [{\citenamefont {McKenna}, \citenamefont {Leterrier},\ and\
  \citenamefont {Schultheisz}(1995)}]{mckenna1995}%
  \BibitemOpen
  \bibfield  {author} {\bibinfo {author} {\bibfnamefont {G.~B.}\ \bibnamefont
  {McKenna}}, \bibinfo {author} {\bibfnamefont {Y.}~\bibnamefont {Leterrier}},\
  and\ \bibinfo {author} {\bibfnamefont {C.~R.}\ \bibnamefont {Schultheisz}},\
  }\bibfield  {title} {\enquote {\bibinfo {title} {The evolution of material
  properties during physical aging},}\ }\href
  {https://doi.org/https://doi.org/10.1002/pen.760350505} {\bibfield  {journal}
  {\bibinfo  {journal} {Polymer Engineering \& Science}\ }\textbf {\bibinfo
  {volume} {35}},\ \bibinfo {pages} {403--410} (\bibinfo {year}
  {1995})}\BibitemShut {NoStop}%
\bibitem [{\citenamefont {McKenna}(2012)}]{mckenna2012}%
  \BibitemOpen
  \bibfield  {author} {\bibinfo {author} {\bibfnamefont {G.~B.}\ \bibnamefont
  {McKenna}},\ }\enquote {\bibinfo {title} {Physical aging in glasses and
  composites},}\ in\ \href {https://doi.org/10.1007/978-1-4419-9308-3_7} {\emph
  {\bibinfo {booktitle} {Long-Term Durability of Polymeric Matrix
  Composites}}},\ \bibinfo {editor} {edited by\ \bibinfo {editor}
  {\bibfnamefont {K.~V.}\ \bibnamefont {Pochiraju}}, \bibinfo {editor}
  {\bibfnamefont {G.~P.}\ \bibnamefont {Tandon}},\ and\ \bibinfo {editor}
  {\bibfnamefont {G.~A.}\ \bibnamefont {Schoeppner}}}\ (\bibinfo  {publisher}
  {Springer US},\ \bibinfo {address} {Boston, MA},\ \bibinfo {year} {2012})\
  pp.\ \bibinfo {pages} {237--309}\BibitemShut {NoStop}%
\bibitem [{\citenamefont {Moynihan}\ \emph
  {et~al.}(1976{\natexlab{b}})\citenamefont {Moynihan}, \citenamefont
  {Easteal}, \citenamefont {DeBolt},\ and\ \citenamefont
  {Tucker}}]{Moynihan1976b}%
  \BibitemOpen
  \bibfield  {author} {\bibinfo {author} {\bibfnamefont {C.~T.}\ \bibnamefont
  {Moynihan}}, \bibinfo {author} {\bibfnamefont {A.~J.}\ \bibnamefont
  {Easteal}}, \bibinfo {author} {\bibfnamefont {M.~A.}\ \bibnamefont
  {DeBolt}},\ and\ \bibinfo {author} {\bibfnamefont {J.}~\bibnamefont
  {Tucker}},\ }\bibfield  {title} {\enquote {\bibinfo {title} {Dependence of
  the fictive temperature of glass on cooling rate},}\ }\href@noop {}
  {\bibfield  {journal} {\bibinfo  {journal} {J. Am. Ceram. Soc.}\ }\textbf
  {\bibinfo {volume} {59}},\ \bibinfo {pages} {12--16} (\bibinfo {year}
  {1976}{\natexlab{b}})}\BibitemShut {NoStop}%
\bibitem [{\citenamefont {Roed}\ \emph {et~al.}(2019)\citenamefont {Roed},
  \citenamefont {Hecksher}, \citenamefont {Dyre},\ and\ \citenamefont
  {Niss}}]{roed2019}%
  \BibitemOpen
  \bibfield  {author} {\bibinfo {author} {\bibfnamefont {L.~A.}\ \bibnamefont
  {Roed}}, \bibinfo {author} {\bibfnamefont {T.}~\bibnamefont {Hecksher}},
  \bibinfo {author} {\bibfnamefont {J.~C.}\ \bibnamefont {Dyre}},\ and\
  \bibinfo {author} {\bibfnamefont {K.}~\bibnamefont {Niss}},\ }\bibfield
  {title} {\enquote {\bibinfo {title} {Generalized single-parameter aging tests
  and their application to glycerol},}\ }\href@noop {} {\bibfield  {journal}
  {\bibinfo  {journal} {J. Chem. Phys.}\ }\textbf {\bibinfo {volume} {150}},\
  \bibinfo {pages} {044501} (\bibinfo {year} {2019})}\BibitemShut {NoStop}%
\bibitem [{\citenamefont {Hecksher}, \citenamefont {Olsen},\ and\ \citenamefont
  {Dyre}(2019)}]{hecksher2019}%
  \BibitemOpen
  \bibfield  {author} {\bibinfo {author} {\bibfnamefont {T.}~\bibnamefont
  {Hecksher}}, \bibinfo {author} {\bibfnamefont {N.~B.}\ \bibnamefont
  {Olsen}},\ and\ \bibinfo {author} {\bibfnamefont {J.~C.}\ \bibnamefont
  {Dyre}},\ }\bibfield  {title} {\enquote {\bibinfo {title} {Fast contribution
  to the activation energy of a glass-forming liquid},}\ }\href@noop {}
  {\bibfield  {journal} {\bibinfo  {journal} {Proc. Natl. Acad. Sci. U. S. A.}\
  }\textbf {\bibinfo {volume} {116}},\ \bibinfo {pages} {16736} (\bibinfo
  {year} {2019})}\BibitemShut {NoStop}%
\bibitem [{\citenamefont {Castillo}\ and\ \citenamefont
  {Parsaeian}(2007)}]{Castillo2007}%
  \BibitemOpen
  \bibfield  {author} {\bibinfo {author} {\bibfnamefont {H.~E.}\ \bibnamefont
  {Castillo}}\ and\ \bibinfo {author} {\bibfnamefont {A.}~\bibnamefont
  {Parsaeian}},\ }\bibfield  {title} {\enquote {\bibinfo {title} {Local
  fluctuations in the ageing of a simple structural glass},}\ }\href
  {https://doi.org/10.1038/nphys482} {\bibfield  {journal} {\bibinfo  {journal}
  {Nature Physics}\ }\textbf {\bibinfo {volume} {3}},\ \bibinfo {pages}
  {26--28} (\bibinfo {year} {2007})}\BibitemShut {NoStop}%
\bibitem [{\citenamefont {Leonardo}\ \emph {et~al.}(2000)\citenamefont
  {Leonardo}, \citenamefont {Angelani}, \citenamefont {Parisi},\ and\
  \citenamefont {Ruocco}}]{DiLeonardo2000}%
  \BibitemOpen
  \bibfield  {author} {\bibinfo {author} {\bibfnamefont {R.~D.}\ \bibnamefont
  {Leonardo}}, \bibinfo {author} {\bibfnamefont {L.}~\bibnamefont {Angelani}},
  \bibinfo {author} {\bibfnamefont {G.}~\bibnamefont {Parisi}},\ and\ \bibinfo
  {author} {\bibfnamefont {G.}~\bibnamefont {Ruocco}},\ }\bibfield  {title}
  {\enquote {\bibinfo {title} {Off-equilibrium effec- tive temperature in
  monatomic lennard-jones glass},}\ }\href@noop {} {\bibfield  {journal}
  {\bibinfo  {journal} {Physical Review Letters}\ }\textbf {\bibinfo {volume}
  {84}},\ \bibinfo {pages} {6054} (\bibinfo {year} {2000})}\BibitemShut
  {NoStop}%
\bibitem [{\citenamefont {Gnan}\ \emph {et~al.}(2010)\citenamefont {Gnan},
  \citenamefont {Maggi}, \citenamefont {Schr\o{}der},\ and\ \citenamefont
  {Dyre}}]{Gnan2010}%
  \BibitemOpen
  \bibfield  {author} {\bibinfo {author} {\bibfnamefont {N.}~\bibnamefont
  {Gnan}}, \bibinfo {author} {\bibfnamefont {C.}~\bibnamefont {Maggi}},
  \bibinfo {author} {\bibfnamefont {T.~B.}\ \bibnamefont {Schr\o{}der}},\ and\
  \bibinfo {author} {\bibfnamefont {J.~C.}\ \bibnamefont {Dyre}},\ }\bibfield
  {title} {\enquote {\bibinfo {title} {Predicting the effective temperature of
  a glass},}\ }\href {https://doi.org/10.1103/PhysRevLett.104.125902}
  {\bibfield  {journal} {\bibinfo  {journal} {Phys. Rev. Lett.}\ }\textbf
  {\bibinfo {volume} {104}},\ \bibinfo {pages} {125902} (\bibinfo {year}
  {2010})}\BibitemShut {NoStop}%
\bibitem [{\citenamefont {Schober}(2012)}]{Schober2012}%
  \BibitemOpen
  \bibfield  {author} {\bibinfo {author} {\bibfnamefont {H.~R.}\ \bibnamefont
  {Schober}},\ }\bibfield  {title} {\enquote {\bibinfo {title} {Modeling aging
  rates in a simple glass and its melt},}\ }\href
  {https://doi.org/10.1103/PhysRevB.85.024204} {\bibfield  {journal} {\bibinfo
  {journal} {Phys. Rev. B}\ }\textbf {\bibinfo {volume} {85}},\ \bibinfo
  {pages} {024204} (\bibinfo {year} {2012})}\BibitemShut {NoStop}%
\bibitem [{\citenamefont {Hecksher}\ \emph {et~al.}(2010)\citenamefont
  {Hecksher}, \citenamefont {Olsen}, \citenamefont {Niss},\ and\ \citenamefont
  {Dyre}}]{hecksher2010}%
  \BibitemOpen
  \bibfield  {author} {\bibinfo {author} {\bibfnamefont {T.}~\bibnamefont
  {Hecksher}}, \bibinfo {author} {\bibfnamefont {N.~B.}\ \bibnamefont {Olsen}},
  \bibinfo {author} {\bibfnamefont {K.}~\bibnamefont {Niss}},\ and\ \bibinfo
  {author} {\bibfnamefont {J.~C.}\ \bibnamefont {Dyre}},\ }\bibfield  {title}
  {\enquote {\bibinfo {title} {Physical aging of molecular glasses studied by a
  device allowing for rapid thermal equilibration},}\ }\href@noop {} {\bibfield
   {journal} {\bibinfo  {journal} {J. Chem. Phys.}\ }\textbf {\bibinfo {volume}
  {133}},\ \bibinfo {pages} {174514} (\bibinfo {year} {2010})}\BibitemShut
  {NoStop}%
\end{thebibliography}
%

\clearpage
\onecolumngrid
\setcounter{section}{0}
\setcounter{equation}{0}
\setcounter{figure}{0}
\setcounter{page}{1}
\renewcommand{\thepage}{S\arabic{page}}
\renewcommand{\theequation}{S\arabic{equation}}
\renewcommand{\thefigure}{S\arabic{figure}}

\section{Single Parameter Aging and Density Scaling: Supplemental Material}

\subsection{Details of the SPA data treatment}
\label{sec:details}

In this section we describe the details of the SPA analysis performed on the temperature-jump relaxation data of Di Lisio \textit{et al} \cite{dilisio2023}. 
The calculation of a non-linear jump from another non-linear jump proceeds via Eq.~(\ref{eq:tint}) of the paper
\begin{equation}
t_2=\frac{\tau_{\text{eq},2}}{\tau_{\text{eq},1}}\int_0^{t_1} \exp \left\{c(\Delta T_1-\Delta T_2)R  \right\}\, dt_1\,,
\end{equation}
where the subscripts 1 and 2 refer to temperature jump 1 and temperature jump 2.

This calculation requires the knowledge of equilibrium relaxation times at the bath temperatures, $\tau_{\text{eq},1}$ and $\tau_{\text{eq},2}$ and the constant $c$.
The constant $c$ can be determined from data from two different temperature jump experiments. We follow the approach described in Refs.~\onlinecite{hecksher2019, niss2020}, where the Kovacs-McKenna (KM) relaxation rate defined as
\begin{equation}
    \Gamma_\text{KM}=-\frac{d\ln R}{dt}
\end{equation}
is used as an intermediate step. 
The dimensionless KM relaxation rate is similarly defined \cite{hecksher2010} with material time replacing time
\begin{equation}
    \tilde{\Gamma}_\text{KM}=-\frac{d\ln R}{d\xi}
\end{equation}
which via the chain rule is equivalent to
\begin{equation}\label{eq:dimGamma}
    \tilde{\Gamma}_\text{KM}=-\frac{d\ln R}{dt}\frac{dt}{d\xi} = \Gamma_\text{KM}\tau\,,
\end{equation}
where the last equality sign comes from the definition of the material time, $d\xi=dt/\tau(t)$. 

Combining Eqs.~(\ref{eq:gamma}) and (\ref{eq:dimGamma}) we end up with 
\begin{equation}\label{eq:dimGamma2}
    \ln \tilde{\Gamma}_\text{KM} = \ln \Gamma_\text{KM} + \ln\tau = \ln\Gamma_\text{KM} + \ln\tau_\text{eq}+c\Delta T R\,,
\end{equation}
thus determining $c$ as the value that collapses $\tilde{\Gamma}_\text{KM}$ for all temperature jumps.

\begin{figure}
    \centering
    \includegraphics[width=13cm]{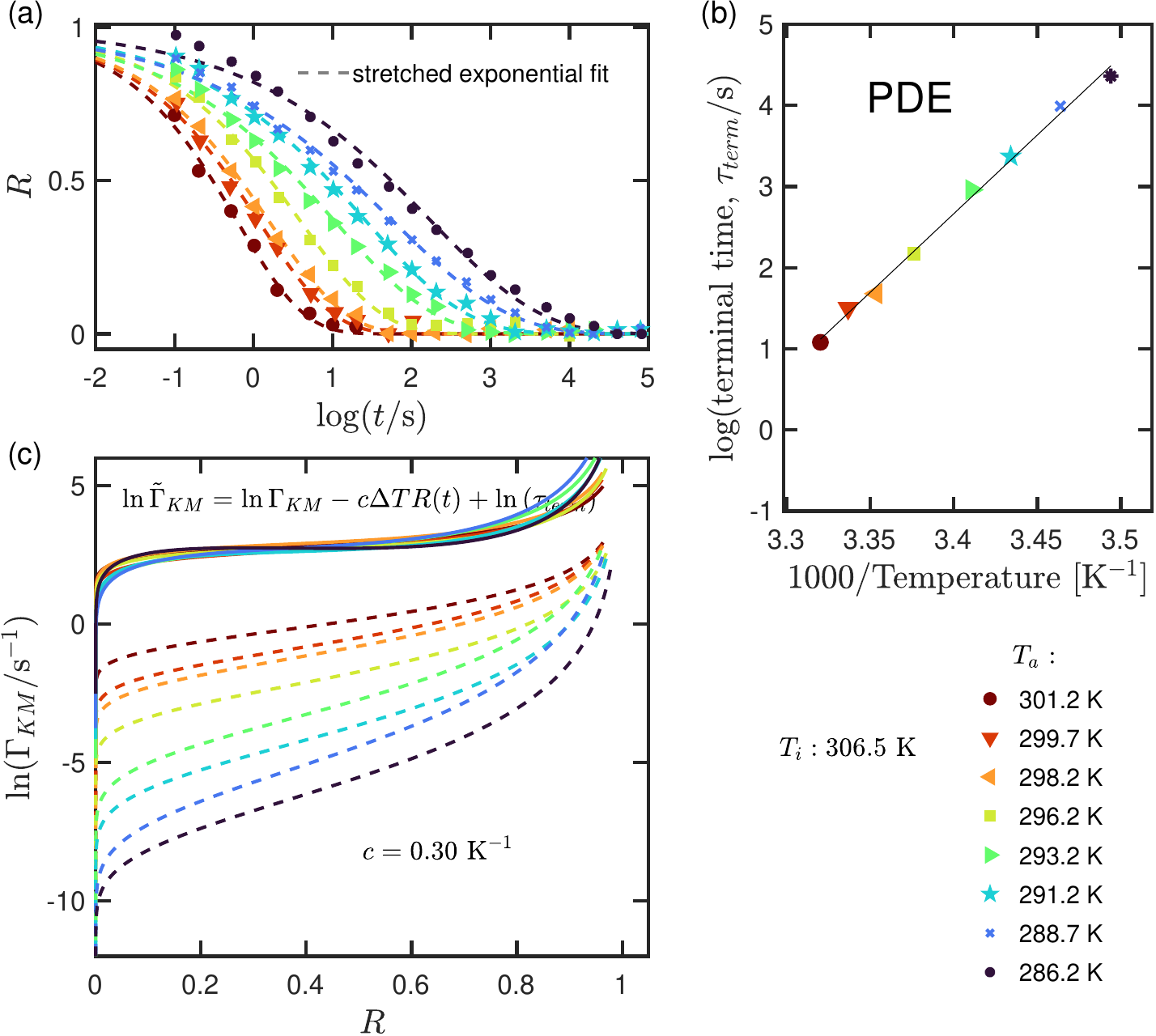}
    \caption{Steps in the SPA analysis of the Di Lisio data. $T_i$ is the initial temperature, $T_a$ is the annealing temperature. (a) Data points and the corresponding stretched exponential fits. (b) The extracted terminal (equilibrium) relaxation time. This time is defined as the terminal time via $R(\tau_\text{term})=0.01$, i.e. as the time where the fitted relaxation curve is 1\% from equilibrium. (c) The KM relaxation rate for the fitted curves in (a) are shown as dashed curves. The dimensionless KM relaxation rates with $c=0.30$~K$^{-1}$ (full lines) are shown to collapse to a good approximation.}
    \label{fig:PDE_determine_c}
\end{figure}

The data points are sparse and somewhat noisy, so the approach involving a derivative is not really feasible without applying a fitting function to interpolate between data points. We thus fitted each relaxation curve to a stretched exponential and calculating $\Gamma_\text{KM}$ from the fits. The equilibrium relaxation time $\tau_\text{eq}$ was also estimated as the terminal time, $\tau_\text{term}$, defined as $R(\tau_\text{term})=0.01$.
Subsequently, the $c$ was optimized for the best collapse of $\Tilde{\Gamma}$ with $\tau_\text{term}$ replacing $\tau_\text{eq}$ in Eq.~(\ref{eq:dimGamma}). 
This procedure strictly forces the predicted curves to match long time tail of the aging curves. However, according to Di Lisio \textit{et al.}\cite{dilisio2023} the terminal times of the aging curves do agree with the structural relaxation times, so this is not imposing any unwarranted restrictions on the predictions.

\begin{figure}[h!]
   \centering
    \includegraphics[width=15cm]{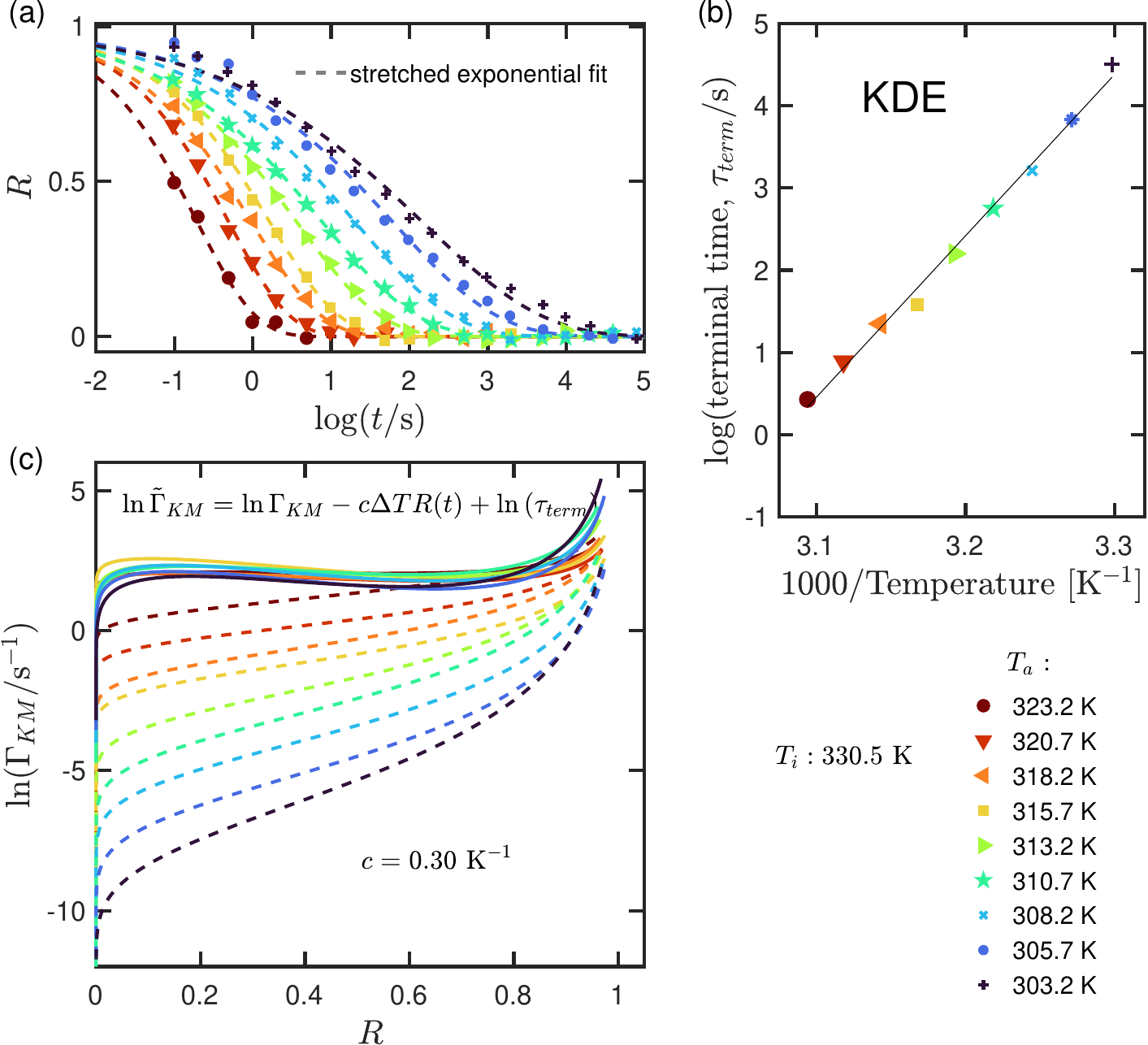}
    \caption{Steps in the SPA analysis for KDE. (a) Data points and the corresponding stretched exponential fits. (b) The extracted terminal (equilibrium) relaxation time. This time is defined as the terminal time via $R(\tau_\text{term})=0.01$, i.e. as the time where the fitted relaxation curve is 1\% from equilibrium. (c) The KM relaxation rate for the fitted curves in (a) are shown as dashed curves. The dimensionless KM relaxation rates with $c=0.30$~K$^{-1}$ (full lines) are shown to collapse to a good approximation.}
    \label{fig:KDE_deta}
\end{figure}

\begin{figure}[h!]
    \centering
    \includegraphics[width=15cm]{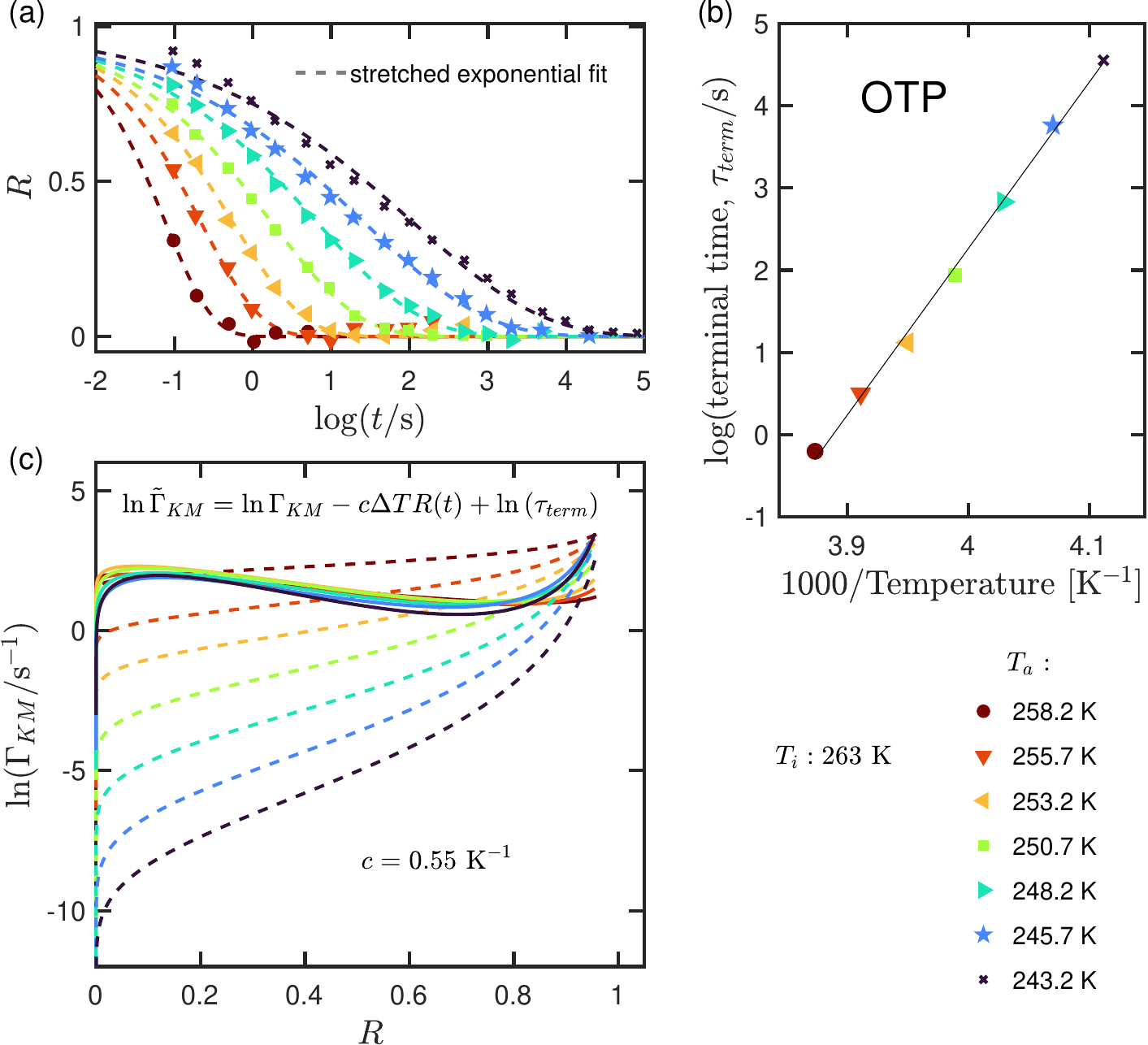}
    \caption{Steps in the SPA analysis for OTP. (a) Data points and the corresponding stretched exponential fits. (b) The extracted terminal (equilibrium) relaxation time. This time is defined as the terminal time via $R(\tau_\text{term})=0.01$, i.e. as the time where the fitted relaxation curve is 1\% from equilibrium. (c) The KM relaxation rate for the fitted curves in (a) are shown as dashed curves. The dimensionless KM relaxation rates with $c=0.55$~K$^{-1}$ (full lines) are shown to collapse to a good approximation.}
    \label{fig:OTP_deta}
\end{figure}

\begin{figure}[h!]
    \centering
    \includegraphics[width=15cm]{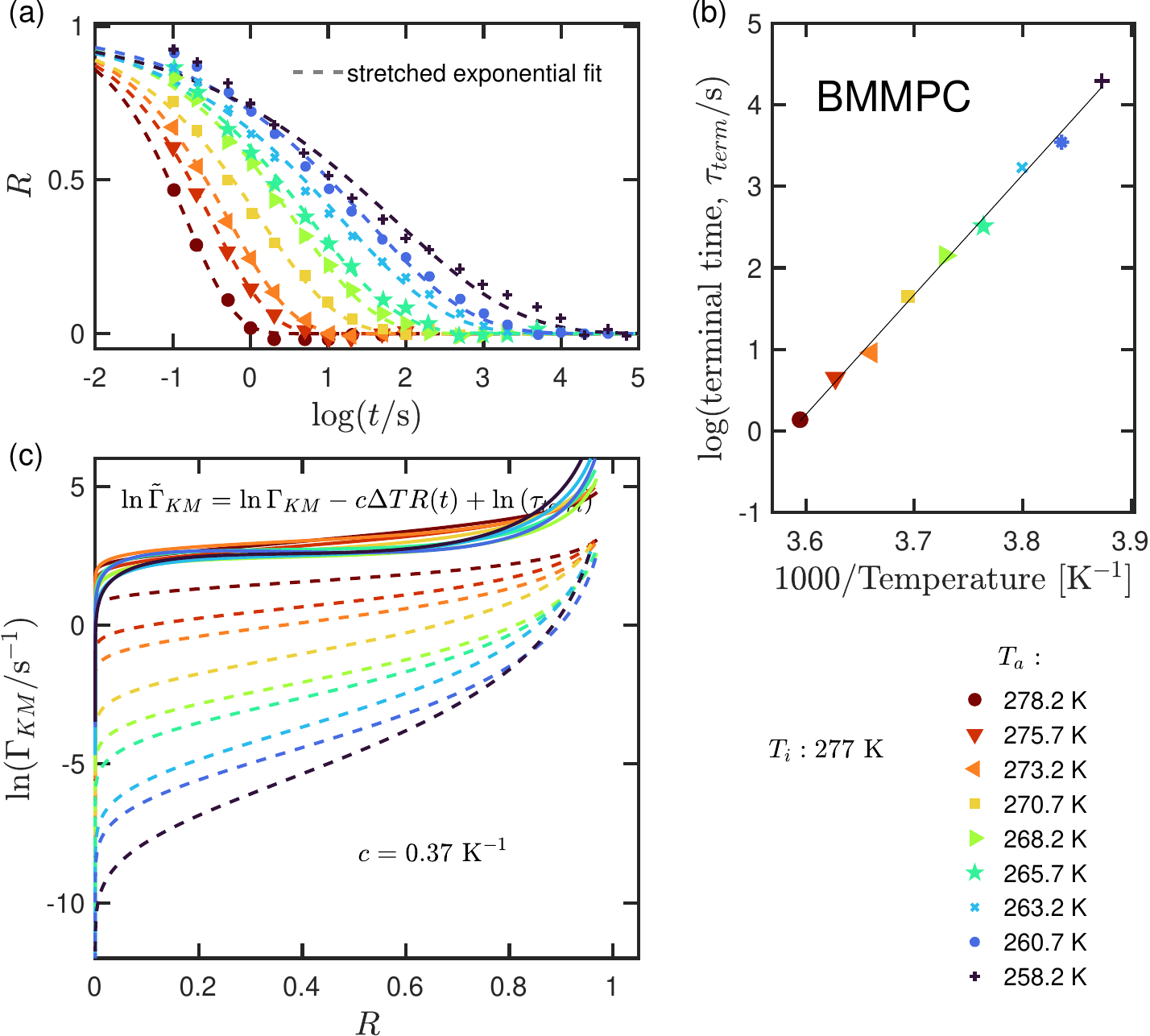}
    \caption{Steps in the SPA analysis for BMMPC. (a) Data points and the corresponding stretched exponential fits. (b) The extracted terminal (equilibrium) relaxation time. This time is defined as the terminal time via $R(\tau_\text{term})=0.01$, i.e. as the time where the fitted relaxation curve is 1\% from equilibrium. (c) The KM relaxation rate for the fitted curves in (a) are shown as dashed curves. The dimensionless KM relaxation rates with $c=0.37$~K$^{-1}$ (full lines) are shown to collapse to a good approximation.}
    \label{fig:BMMPC_deta}
\end{figure}

\begin{figure}[h!]
    \centering
    \includegraphics[width=15cm]{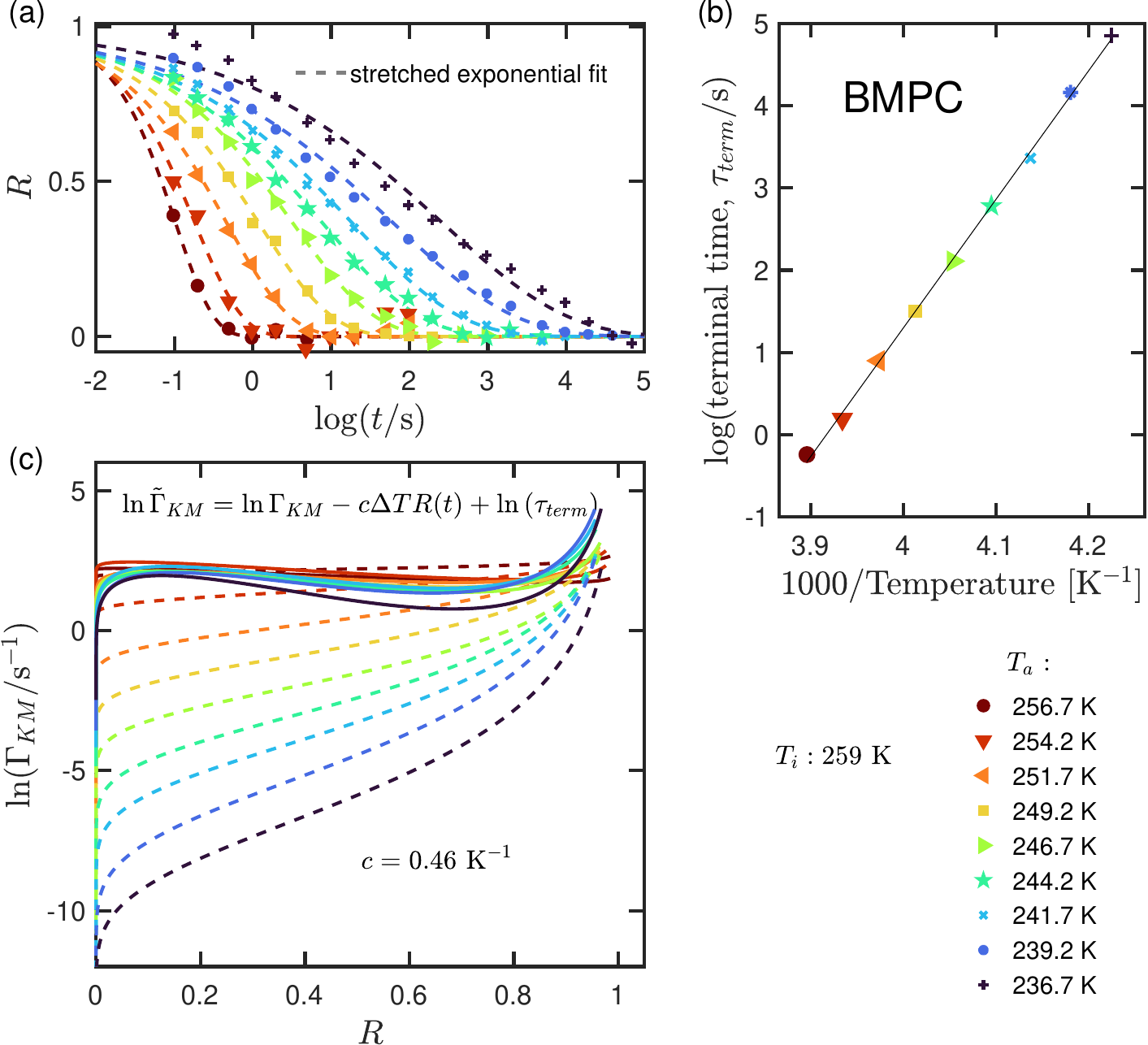}
    \caption{Steps in the SPA analysis for BMPC. (a) Data points and the corresponding stretched exponential fits. (b) The extracted terminal (equilibrium) relaxation time. This time is defined as the terminal time via $R(\tau_\text{term})=0.01$, i.e. as the time where the fitted relaxation curve is 1\% from equilibrium. (c) The KM relaxation rate for the fitted curves in (a) are shown as dashed curves. The dimensionless KM relaxation rates with $c=0.46$~K$^{-1}$ (full lines) are shown to collapse to a good approximation.}
    \label{fig:BMPC_deta}
\end{figure}

Figures \ref{fig:PDE_determine_c}-\ref{fig:BMPC_deta} illustrate the steps in this procedure for each of the studied substances. Figure \ref{fig:PDE_determine_c}(A) shows the data points and the corresponding stretched exponential fits as dashed lines. Figure \ref{fig:PDE_determine_c}(B) shows the extracted terminal time scales vs inverse temperature with an Arrhenius fit as a line. There is some scatter in the values which comes partly comes from scatter in the relaxation data and partly from imperfect fits. Finally, Fig.~\ref{fig:PDE_determine_c}(C) shows the KM relaxation rate as dashed lines and the collapse of the dimensionless KM relaxation rate with $c=-0.30$~K$^{-1}$. This value for $c$ can be used to calculate from one temperature jump to any other temperature jump -- given that SPA applies.

The collapse in $\tilde{\Gamma}_\text{KM}$ is not perfect for all the liquids, but one has to keep in mind that the curves are fits to the data points and especially the large down-jumps are less well fitted by a stretched exponential. In fact, the predictions in Fig.~\ref{fig:org_spa} capture the data points better than the stretched exponential fit.


\begin{figure}
    \centering
    \includegraphics[width=1\linewidth]{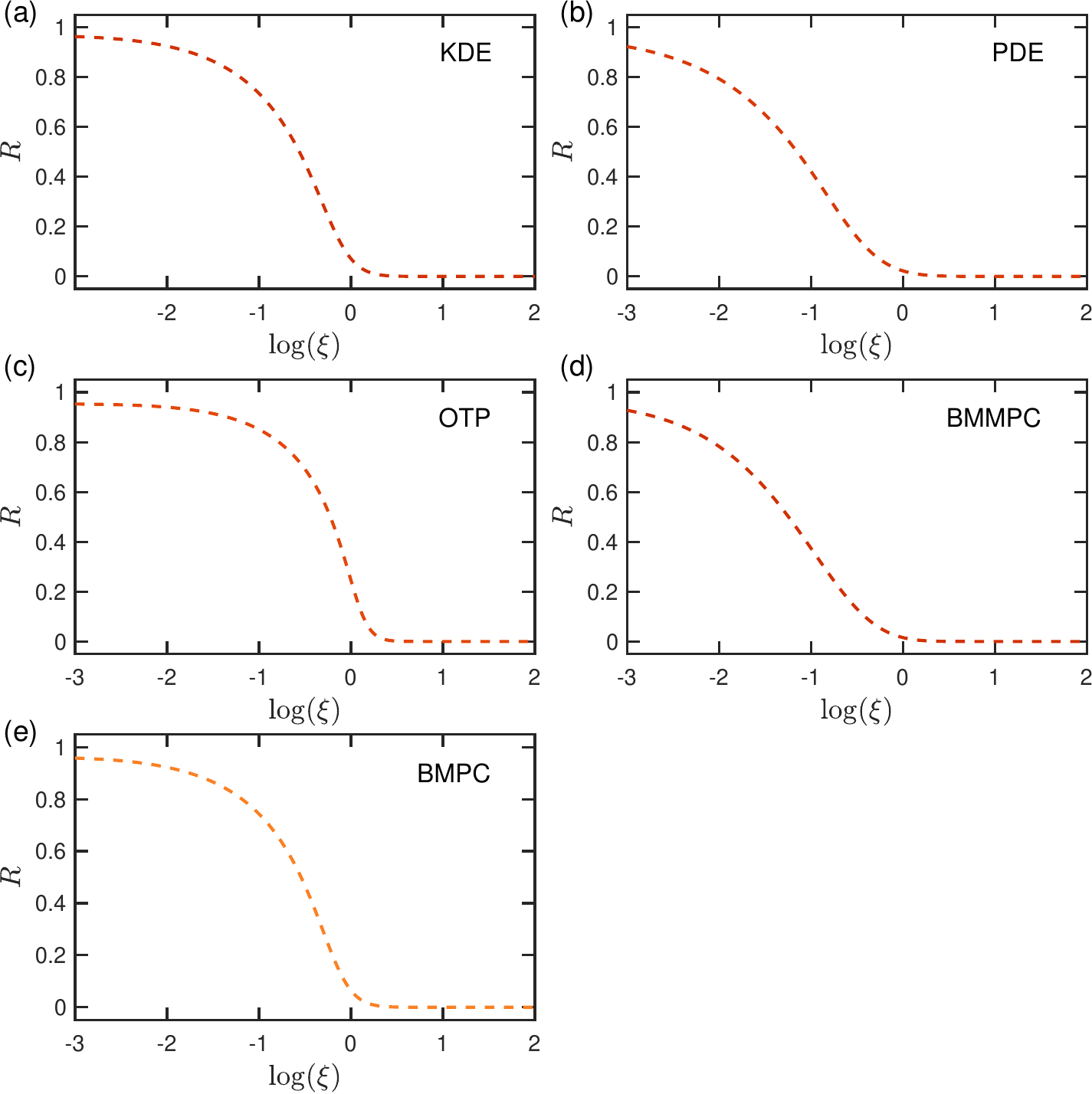}
    \caption{The linear aging curves. In the paper, the stretched exponential fit to one set of data is used to predict the other temperature jumps. The chosen curve can also be used to predict the linear jump (in TN formalism this corresponds to the memory kernel of the linear response integral). This is simply obtained by setting $\Delta T_2=0$ in Eq. (6) of the paper. To obtain the matertial time, $\xi$, we have normalised the time axis by the equilibrium relaxation time.}
    \label{fig:enter-label}
\end{figure}

\end{document}